\documentclass[preprint2,trackchanges]{aastex631}%This is for two column

\usepackage{amsmath,multirow, breakurl,hyperref}
\usepackage[T1]{fontenc}

\newcommand\aastex{AAS\TeX}

\shorttitle{\aastex\ MWL Observations of VER~J0521+211 In an Elevated $\gamma$-Ray State}
\shortauthors{VERITAS et al.}

\begin{document}

\title{Multiwavelength Observations of the Blazar VER~J0521+211 During an Elevated TeV Gamma-Ray State}

\correspondingauthor{Qi Feng}
\email{qifeng@nevis.columbia.edu}
\correspondingauthor{Cornelia Arcaro}
\email{contact.magic@mpp.mpg.de}
\suppressAffiliations
\author{C.~B.~Adams}\affiliation{Physics Department, Columbia University, New York, NY 10027, USA}
\author{P.~Batista}\affiliation{DESY, Platanenallee 6, 15738 Zeuthen, Germany}
\author{W.~Benbow}\affiliation{Center for Astrophysics $|$ Harvard \& Smithsonian, Cambridge, MA 02138, USA}
\author{A.~Brill}\affiliation{N.A.S.A./Goddard Space-Flight Center, Code 661, Greenbelt, MD 20771, USA}
\author{R.~Brose}\affiliation{Institute of Physics and Astronomy, University of Potsdam, 14476 Potsdam-Golm, Germany and DESY, Platanenallee 6, 15738 Zeuthen, Germany}
\author{J.~H.~Buckley}\affiliation{Department of Physics, Washington University, St. Louis, MO 63130, USA}
\author{M.~Capasso}\affiliation{Department of Physics and Astronomy, Barnard College, Columbia University, NY 10027, USA}
\author{J.~L.~Christiansen}\affiliation{Physics Department, California Polytechnic State University, San Luis Obispo, CA 94307, USA}
\author{M.~Errando}\affiliation{Department of Physics, Washington University, St. Louis, MO 63130, USA}
\author{Q.~Feng}\affiliation{Department of Physics and Astronomy, Barnard College, Columbia University, NY 10027, USA}
\author{J.~P.~Finley}\affiliation{Department of Physics and Astronomy, Purdue University, West Lafayette, IN 47907, USA}
\author{G.~M.~Foote}\affiliation{Department of Physics and Astronomy and the Bartol Research Institute, University of Delaware, Newark, DE 19716, USA}
\author{L.~Fortson}\affiliation{School of Physics and Astronomy, University of Minnesota, Minneapolis, MN 55455, USA}
\author{A.~Furniss}\affiliation{Department of Physics, California State University - East Bay, Hayward, CA 94542, USA}
\author{G.~Gallagher}\affiliation{Department of Physics and Astronomy, Ball State University, Muncie, IN 47306, USA}
\author{A.~Gent}\affiliation{School of Physics and Center for Relativistic Astrophysics, Georgia Institute of Technology, 837 State Street NW, Atlanta, GA 30332-0430}
\author{C.~Giuri}\affiliation{DESY, Platanenallee 6, 15738 Zeuthen, Germany}
\author{W.~F~Hanlon}\affiliation{Center for Astrophysics $|$ Harvard \& Smithsonian, Cambridge, MA 02138, USA}
\author{D.~Hanna}\affiliation{Physics Department, McGill University, Montreal, QC H3A 2T8, Canada}
\author{T.~Hassan}\affiliation{DESY, Platanenallee 6, 15738 Zeuthen, Germany}
\author{O.~Hervet}\affiliation{Santa Cruz Institute for Particle Physics and Department of Physics, University of California, Santa Cruz, CA 95064, USA}
\author{J.~Holder}\affiliation{Department of Physics and Astronomy and the Bartol Research Institute, University of Delaware, Newark, DE 19716, USA}
\author{B.~Hona}\affiliation{Department of Physics and Astronomy, University of Utah, Salt Lake City, UT 84112, USA}
\author{G.~Hughes}\affiliation{Center for Astrophysics $|$ Harvard \& Smithsonian, Cambridge, MA 02138, USA}
\author{T.~B.~Humensky}\affiliation{Physics Department, Columbia University, New York, NY 10027, USA}
\author{W.~Jin}\affiliation{Department of Physics and Astronomy, University of Alabama, Tuscaloosa, AL 35487, USA}
\author{P.~Kaaret}\affiliation{Department of Physics and Astronomy, University of Iowa, Van Allen Hall, Iowa City, IA 52242, USA}
\author{M.~Kertzman}\affiliation{Department of Physics and Astronomy, DePauw University, Greencastle, IN 46135-0037, USA}
\author{D.~Kieda}\affiliation{Department of Physics and Astronomy, University of Utah, Salt Lake City, UT 84112, USA}
\author{T.~K~Kleiner}\affiliation{DESY, Platanenallee 6, 15738 Zeuthen, Germany}
\author{F.~Krennrich}\affiliation{Department of Physics and Astronomy, Iowa State University, Ames, IA 50011, USA}
\author{S.~Kumar}\affiliation{Physics Department, McGill University, Montreal, QC H3A 2T8, Canada}
\author{M.~J.~Lang}\affiliation{School of Physics, National University of Ireland Galway, University Road, Galway, Ireland}
\author{M.~Lundy}\affiliation{Physics Department, McGill University, Montreal, QC H3A 2T8, Canada}
\author{G.~Maier}\affiliation{DESY, Platanenallee 6, 15738 Zeuthen, Germany}
\author{J.~Millis}\affiliation{Department of Physics and Astronomy, Ball State University, Muncie, IN 47306, USA and Department of Physics, Anderson University, 1100 East 5th Street, Anderson, IN 46012}
\author{P.~Moriarty}\affiliation{School of Physics, National University of Ireland Galway, University Road, Galway, Ireland}
\author{R.~Mukherjee}\affiliation{Department of Physics and Astronomy, Barnard College, Columbia University, NY 10027, USA}
\author{M.~Nievas-Rosillo}\affiliation{DESY, Platanenallee 6, 15738 Zeuthen, Germany}
\author{S.~O'Brien}\affiliation{Physics Department, McGill University, Montreal, QC H3A 2T8, Canada}
\author{R.~A.~Ong}\affiliation{Department of Physics and Astronomy, University of California, Los Angeles, CA 90095, USA}
\author{A.~N.~Otte}\affiliation{School of Physics and Center for Relativistic Astrophysics, Georgia Institute of Technology, 837 State Street NW, Atlanta, GA 30332-0430}
\author{S.~Patel}\affiliation{Department of Physics and Astronomy, University of Iowa, Van Allen Hall, Iowa City, IA 52242, USA}
\author{S.~R.~Patel}\affiliation{DESY, Platanenallee 6, 15738 Zeuthen, Germany}
\author{K.~Pfrang}\affiliation{DESY, Platanenallee 6, 15738 Zeuthen, Germany}
\author{M.~Pohl}\affiliation{Institute of Physics and Astronomy, University of Potsdam, 14476 Potsdam-Golm, Germany and DESY, Platanenallee 6, 15738 Zeuthen, Germany}
\author{R.~R.~Prado}\affiliation{DESY, Platanenallee 6, 15738 Zeuthen, Germany}
\author{E.~Pueschel}\affiliation{DESY, Platanenallee 6, 15738 Zeuthen, Germany}
\author{J.~Quinn}\affiliation{School of Physics, University College Dublin, Belfield, Dublin 4, Ireland}
\author{K.~Ragan}\affiliation{Physics Department, McGill University, Montreal, QC H3A 2T8, Canada}
\author{P.~T.~Reynolds}\affiliation{Department of Physical Sciences, Munster Technological University, Bishopstown, Cork, T12 P928, Ireland}
\author{D.~Ribeiro}\affiliation{Physics Department, Columbia University, New York, NY 10027, USA}
\author{E.~Roache}\affiliation{Center for Astrophysics $|$ Harvard \& Smithsonian, Cambridge, MA 02138, USA}
\author{J.~L.~Ryan}\affiliation{Department of Physics and Astronomy, University of California, Los Angeles, CA 90095, USA}
\author{I.~Sadeh}\affiliation{DESY, Platanenallee 6, 15738 Zeuthen, Germany}
\author{M.~Santander}\affiliation{Department of Physics and Astronomy, University of Alabama, Tuscaloosa, AL 35487, USA}
\author{G.~H.~Sembroski}\affiliation{Department of Physics and Astronomy, Purdue University, West Lafayette, IN 47907, USA}
\author{R.~Shang}\affiliation{Department of Physics and Astronomy, University of California, Los Angeles, CA 90095, USA}
\author{B.~Stevenson}\affiliation{Department of Physics and Astronomy, University of California, Los Angeles, CA 90095, USA}
\author{J.~V.~Tucci}\affiliation{Department of Physics, Indiana University-Purdue University Indianapolis, Indianapolis, IN 46202, USA}
\author{V.~V.~Vassiliev}\affiliation{Department of Physics and Astronomy, University of California, Los Angeles, CA 90095, USA}
\author{S.~P.~Wakely}\affiliation{Enrico Fermi Institute, University of Chicago, Chicago, IL 60637, USA}
\author{A.~Weinstein}\affiliation{Department of Physics and Astronomy, Iowa State University, Ames, IA 50011, USA}
\author{R.~M.~Wells}\affiliation{Department of Physics and Astronomy, Iowa State University, Ames, IA 50011, USA}
\author{D.~A.~Williams}\affiliation{Santa Cruz Institute for Particle Physics and Department of Physics, University of California, Santa Cruz, CA 95064, USA}
\author{T.~J~Williamson}\affiliation{Department of Physics and Astronomy and the Bartol Research Institute, University of Delaware, Newark, DE 19716, USA}
\collaboration{0}{The VERITAS Collaboration}
% authors 03.12.2021  Format ApJ
%
%\usepackage[T1]{fontenc}
\author[0000-0001-8307-2007]{V.~A.~Acciari}
\affiliation{Instituto de Astrof\'isica de Canarias and Dpto. de  Astrof\'isica, Universidad de La Laguna, E-38200, La Laguna, Tenerife, Spain}
\author{T.~Aniello}
\affiliation{National Institute for Astrophysics (INAF), I-00136 Rome, Italy}
\author[0000-0002-5613-7693]{S.~Ansoldi}
\affiliation{Universit\`a di Udine and INFN Trieste, I-33100 Udine, Italy}\affiliation{also at International Center for Relativistic Astrophysics (ICRA), Rome, Italy}
\author[0000-0002-5037-9034]{L.~A.~Antonelli}
\affiliation{National Institute for Astrophysics (INAF), I-00136 Rome, Italy}
\author[0000-0001-9076-9582]{A.~Arbet Engels}
\affiliation{Max-Planck-Institut f\"ur Physik, D-80805 M\"unchen, Germany}
\author{C.~Arcaro}
\affiliation{Universit\`a di Padova and INFN, I-35131 Padova, Italy}
\author[0000-0002-4899-8127]{M.~Artero}
\affiliation{Institut de F\'isica d'Altes Energies (IFAE), The Barcelona Institute of Science and Technology (BIST), E-08193 Bellaterra (Barcelona), Spain}
\author[0000-0001-9064-160X]{K.~Asano}
\affiliation{Japanese MAGIC Group: Institute for Cosmic Ray Research (ICRR), The University of Tokyo, Kashiwa, 277-8582 Chiba, Japan}
\author[0000-0002-2311-4460]{D.~Baack}
\affiliation{Technische Universit\"at Dortmund, D-44221 Dortmund, Germany}
\author[0000-0002-1444-5604]{A.~Babi\'c}
\affiliation{Croatian MAGIC Group: University of Zagreb, Faculty of Electrical Engineering and Computing (FER), 10000 Zagreb, Croatia}
% ...
%\author{}
\collaboration{0}{The MAGIC Collaboration}

\begin{abstract}
We report on a long-lasting, elevated gamma-ray flux state from VER~J0521+211 observed by VERITAS, MAGIC, and {\it Fermi}-LAT in 2013 and 2014. 
The peak integral flux above 200 GeV measured with the nightly-binned light curve is $(8.8 \pm 0.4) \times 10^{-7} \;\text{ph}\;\text{m}^{-2}\; \text{s}^{-1}$, or $\sim$37\% of the Crab Nebula flux. 
Multiwavelength observations from X-ray, UV, and optical instruments are also presented. 
A moderate correlation between the X-ray and TeV gamma-ray fluxes was observed, and the X-ray spectrum appeared harder when the flux was higher. 
Using the gamma-ray spectrum and four models of the extragalactic background light (EBL), a conservative 95\% confidence upper limit on the redshift of the source was found to be $z\le0.31$. 
Unlike the gamma-ray and X-ray bands, the optical flux did not increase significantly during the studied period compared to the archival low-state flux. 
The spectral variability from optical to X-ray bands suggests that the synchrotron peak of the spectral energy distribution (SED) may become broader during flaring states, which 
can be adequately described with a one-zone synchrotron self-Compton model 
varying the high-energy end of the underlying particle spectrum. 
The synchrotron peak frequency of the SED, as well as the radio morphology of the jet from the MOJAVE program, are consistent with the source being an intermediate-frequency-peaked BL Lac object.

\end{abstract}

%%%%%%%%%
% Intro
%%%%%%%%%
\section{Introduction} \label{sec:intro}

Blazars are divided into subclasses: high-frequency-peaked BL Lac objects (HBLs), intermediate-frequency-peaked BL Lac objects (IBLs), low-frequency-peaked BL Lac objects (LBLs), and flat spectrum radio quasars (FSRQs), based on the peak frequency of their spectral energy distribution (SED) and the equivalent width of optical emission lines \citep[e.g.,][]{Urry95}. 
TeV gamma rays have been detected from all subclasses of blazars, predominantly from HBLs. A few IBLs/LBLs have also been detected in the TeV gamma-ray band, most frequently during flaring states \citep[e.g.,][]{Arlen13, MAGIC18S50716}, including VER~J0521+211 \citep{Archambault13}. 

Interesting differences in radio morphology between subclasses of blazars, beyond the well-known ``blazar sequence'' \citep{Fossati98}, have been recently proposed \citep[e.g.,][]{Kharb08, Hervet16}. 
Particularly, some blazars, typically belonging to the IBL/LBL subclass, exhibit a combination of stationary  knots and superluminal knots in their jets as observed with the Very Long Baseline Array (VLBA).
Gamma-ray flares contemporaneous with the ejection of superluminal knots have been observed in a few TeV IBLs \citep[e.g.,][]{Arlen13, Abeysekara18, MAGIC18S50716}, providing tantalizing evidence for recollimation shocks \citep[e.g.,][]{Komissarov98,Narayan09}.

The discovery of VER~J0521+211 in the TeV gamma-ray band was made from VERITAS observations of the radio and X-ray source RGB~J0521.8+2112, motivated by a cluster of photons with energies $>$30~GeV in {\it Fermi}-LAT data \citep{Archambault13}. The positions of the source determined with X-ray, GeV and TeV gamma-ray data are all within 0.1$^\circ$. The spatial association across the electromagnetic spectrum and correlated variability, most prominent between X-ray and gamma-ray bands, strongly support the association between VER~J0521+211 and RGB~J0521.8+2112. 

As is the case with many other BL Lac objects, the lack of optical emission features leads to a redshift uncertainty of the source. 
A redshift of $z=0.108$ was reported by \citet{Shaw13}, but later studies were unable to confirm it \citep{Archambault13, Paiano17}. Instead, a lower limit of $z>0.18$ \citep{Paiano17} and an upper limit of  $z<0.34$ \citep{Archambault13} were reported. 

VER~J0521+211 is classified as an IBL as the peak of its spectral energy distribution in a low state lies between $10^{14}$ Hz and $10^{15}$ Hz. However, the source exhibits a harder X-ray spectrum and HBL-like behavior during flares. \citep{Archambault13}. 
In HBLs, it is particularly interesting to study the flux correlation between X-ray and TeV gamma-ray bands, where the SED peaks lie. Despite being model-dependent, the quantitative description of the correlation between X-rays and TeV gamma rays can be informative of the radiative processes of the relativistic particles \citep[e.g.][]{Katarzynski05}. For example, a particle injection in a one-zone synchrotron self-Compton (SSC) model in the Thomson regime would lead to a quadratic correlation between the gamma-ray and X-ray flux, but could lead to a linear correlation if the inverse-Compton (IC) scattering happens in the deep Klein-Nishina (KN) regime \citep[e.g.][]{Aleksic15M42010,Balokovic16}.  

In this work, we report on the results from multiwavelength (MWL) observations of the blazar VER~J0521+211 between 2012 Nov and 2014 Feb, focusing on the TeV gamma-ray and X-ray behaviors. 

%%%%%%%%%
% Observations and data analysis
%%%%%%%%%
\section{Observations and Data Analysis}
\label{sec:obs_data}
%
%%%%%%%%%
\subsection{VERITAS}
%%%%%%%%%
%
The Very Energetic Radiation Imaging Telescope Array System (VERITAS) is an array of four imaging atmospheric Cherenkov telescopes (IACTs) located in southern Arizona \citep[30$^\circ$40'N 110$^\circ$57'W, 1.3 km a.s.l.;][]{Park15}. It is sensitive to gamma rays in the energy range from 85 GeV to $>$30 TeV with an energy resolution of $\sim$15\% (at 1~TeV). The angular resolution (68\% containment at 1~TeV) is $\sim$0.1$^{\circ}$. VERITAS is capable of making a detection at a statistical significance of 5 standard deviations (5 $\sigma$) of a point source of 1\% Crab Nebula flux in $\sim$25~hours, with an energy threshold of 240 GeV when a set of a priori data selection cuts optimized on sources with a moderate power-law index (from $-2.5$ to $-3$) is used.

After its initial discovery in the TeV band \citep{Archambault13}, VER J0521+211 was observed again by VERITAS from 2012 Nov to 2014 Feb, with a total exposure of 23.6 h after data quality selection and dead time correction. 
We analyzed the data using two independent analysis packages \citep[][]{Cogan08, Maier17}, with 
{\it a priori} cuts optimized for lower-energy showers \citep[see e.g.][]{Archambault14}. 
The results of both analysis are compatible and here we use the analysis package described in \citet[][]{Cogan08}.

%%%%%%%%%
\subsection{MAGIC}
%%%%%%%%%
MAGIC is an array of two IACTs located on the Canary Island of La Palma \citep[28$^\circ$45'43''N 17$^\circ$53'24''W, 2.2 km a.s.l.;][]{Aleksic16}. 
MAGIC is capable of detecting a point source of $\sim0.7\%$ of the Crab Nebula flux above 220 GeV in $\sim50$~hours with a statistical significance of 5 $\sigma$ \citep{magicper}. Above 220 GeV, its energy resolution is $\sim16$\% and its angular resolution (68\% containment) is $\lesssim0.07^{\circ}$. 

VER~J0521+211 was observed by MAGIC on four nights between 2013 Oct and 2013 Dec, triggered by elevated optical and GeV gamma-ray fluxes \citep{Buson2013ATel}. 
After data quality selection and dead time correction, a total effective time of $\sim4.5$ hours was taken on 2013 Oct 15, Oct 16, Nov 29, and Dec 2. Details on the stereo analysis of the MAGIC data can be found in \citet{Aleksic11}. \\

%%%%%%%%%
\subsection{{\it Fermi}-LAT}
%%%%%%%%%
The Large Area Telescope (LAT) on-board the {\it Fermi} satellite is a pair-conversion gamma-ray telescope sensitive to energies from $\sim$20~MeV to $>$300~GeV \citep{Atwood09}. 

We performed an unbinned likelihood analysis covering the period of the TeV gamma-ray observations using the LAT ScienceTools \texttt{v11r5p3} and Pass-8 \texttt{P8R2\_SOURCE\_V6\_v06} instrument response functions \citep{Atwood13}. We selected \texttt{SOURCE} class events with an energy between 100 MeV and 300 GeV within 10$^\circ$ from the direction of VER~J0521+211, with a maximum zenith-angle cut of 90$^\circ$. 
We included all point sources within 20$^\circ$ from the source direction in our model, while only leaving the normalization parameters free for those within 5$^\circ$ from the source, and fixed the parameters to the \textit{Fermi} Large Area Telescope Third Source Catalog (3FGL; \citealt{3fgl}) values for those further away. We also included the 3FGL Galactic (\texttt{gll\_iem\_v06}) and isotropic (\texttt{iso\_P8R2\_SOURCE\_V6\_v06}) diffuse emission in the model.

%%%%%%%%%
\subsection{{\it Swift}-XRT}
\label{subsec:XRT}
%%%%%%%%%
The X-Ray Telescope (XRT) on the Neil Gehrels {{\it Swift}} Observatory is a grazing-incidence focusing X-ray telescope, sensitive to energies from 0.2~keV to 10~keV \citep{Gehrels04, Burrows05}. There were 32 \textit{Swift}-XRT observations taken on VER~J0521+211 in the period of interest, 28 of which were taken in photon-counting (PC) mode between 2013 Oct and 2014 Feb, and four in window-timing (WT) mode in 2012 Nov. 
We processed these XRT data using \texttt{xrtpipeline}.\footnote{\texttt{HEASOFT v6.23}, \texttt{swxrtdas\_23Jan18\_v3.4.1} with calibrations from the database \texttt{CALDB 20171113}.} We selected photons with an energy between 0.3 keV and 10 keV and within a circular source region of radius 20 pixels ($\sim47.2''$) to estimate the flux of the source; while those within an annular background region with inner and outer radii of 70 and 120 pixels ($\sim2.75'-4.72'$) were selected to estimate the background contribution. 
For those observations with a count rate in the source region above 0.5 cts s$^{-1}$, we checked that the point spread function is well described by the King function \citep{Moretti05}, and therefore it is not necessary to correct for the pile-up effect. 

We used an absorbed log-parabola model 
to fit the X-ray spectrum: 
\begin{equation}
\label{eqXspec}
\frac{dN}{dE} = e^{-N_H\sigma(E)} K \left( \frac{E}{1\;\text{keV}} \right)^{-\alpha-\beta\log(E/1\;\text{keV})}, 
\end{equation}
where in the absorption component, $N_H$ represents the column density of neutral hydrogen and $\sigma(E)$ is the photoelectric cross-section, while in the power-law component $K$ is the normalization, $\alpha$ is the photon index at 1~keV, and $\beta$ describes the spectral curvature, i.e., the energy dependence of the photon index. 
We fixed the value of $N_H$ to the total neutral hydrogen density, $N_H = 4.38 \times10^{21} \;\text{cm}^{-2}$, derived from the Leiden/Argentine/Bonn (LAB) survey of Galactic HI \citep{Kalberla05}, following \citet{Willingale13}. 
%

%%%%%%%%%
\subsection{{\it Swift}-UVOT}
\label{subsec:UVOT}
%%%%%%%%%

The {\it Swift}-Ultraviolet/Optical telescope (UVOT; \citealt{2005SSRv..120...95R}) was used in some of the observations of VER~J0521+211 during the time period of interest.
The counts from the source were extracted from an aperture of a radius of $5.0''$ centered on the position of the source, 
and the background counts were estimated using the counts from four neighboring regions without any bright source, each having the same radius. The magnitude of the source was then computed using \texttt{uvotsource},\footnote{\texttt{HEASOFT v6.23} with calibrations from the database \texttt{CALDB 20170922}.} and later converted to flux using the zero-point for each of the UVOT filters (the central wavelengths of which are: $V$ 5468 \AA, $B$ 4392 \AA, $U$ 3465 \AA, $UVW1$ 2600 \AA, $UVM2$ 2246 \AA, and $UVW2$ 1928 \AA) from \citet{Poole08}.  

An extinction correction was applied following the procedure described in \citet{2009ApJ...690..163R}, using the value of $E(B-V)$=0.605 \citep{SandF2011}. 
Since VER~J0521+211 is located close to the Galactic plane, with a Galactic latitude of -8.7$^\circ$, the uncertainty on the dust extinction is large. A reddening value of $E(B-V)$=0.703 was found for the direction towards VER~J0521+211 using 472 elliptical galaxies to calibrate infrared dust maps \citep{SFD1998}. The value $E(B-V)$=0.605 used in this work was derived more recently using 261496 stars from the Sloan Digital Sky Survey \citep{SandF2011}, which is 14\% lower than the value from \citet{SFD1998}. \citet{SandF2011} pointed out that if substantial dust exists beyond a distance of 6 kpc, the reddening and extinction would be underestimated.

%%%%%%%%%
\subsection{Steward Observatory}
%%%%%%%%%

VER~J0521+211 was monitored in the $V$-band by the Steward Observatory \citep{Smith09}. There were 76 observations on 37 nights between 2013 Nov 25 and 2014 Apr 1. We use the $V$-band polarimetry results from that are publicly available \footnote{\href{http://james.as.arizona.edu/~psmith/Fermi}{http://james.as.arizona.edu/$\sim$psmith/Fermi}} in this work.

\begin{figure*}[ht!]
\hspace{-0.5cm}
\fig{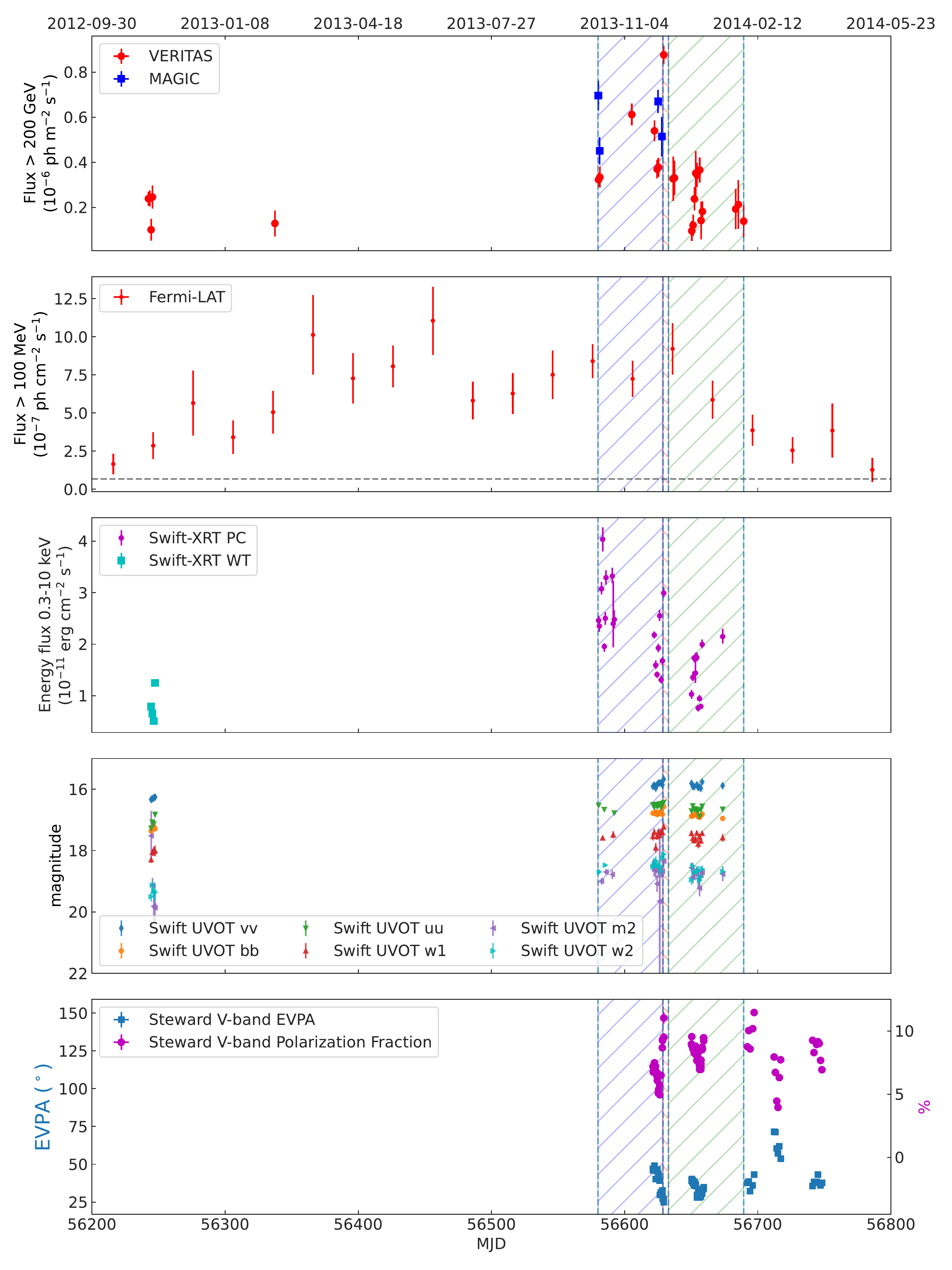}{0.9\textwidth}{}
\caption{MWL light curves of VER~J0521+211 of the entire period of interest in this work, from late 2012 to early 2014. The vertical dashed lines and the colored hatch fills illustrate the three Bayesian blocks between late 2013 and early 2014, based on which we selected the three periods for SED modeling. The binning of the light curves is nightly for VERITAS and MAGIC, monthly for {\it Fermi}-LAT, and by observations for {\it Swift} and Steward Observatory. Only 1-$\sigma$ statistical uncertainties are shown. 
\label{fig:lc}}
\end{figure*}

%%%%%%%%%
% Results
%%%%%%%%%
\section{Results}
\label{sec:res}
%
%%%%%%%%%
\subsection{Temporal Variability}
%
%%%%%%%%%
\subsubsection{Gamma-Ray Light Curves} 
\label{subsec:GRLC}

From the VERITAS observations between 2012 Nov and 2013 Feb, which yielded a quality-selected live time of 4.6 h, the source was detected at a statistical significance of $18 \sigma$ at a time-averaged integral flux above 200~GeV of $(2.4 \pm 0.2) \times 10^{-7} \;\text{photon} \;\text{m}^{-2}\; \text{s}^{-1}$. 
From the 16.4 hours of observations between 2013 Oct and 2014 Feb, the source was detected at $62 \sigma$ and $(3.9 \pm 0.1) \times 10^{-7} \;\text{photon} \;\text{m}^{-2}\; \text{s}^{-1}$ above 200 GeV. 

An overall statistical significance of 30.5 $\sigma$ was obtained from the analysis of the MAGIC observations on the four nights during the period of interest. The energy threshold of these observations was estimated to be $\sim65$ GeV. 
The average flux above 200 GeV of these four MAGIC observations is $(5.8 \pm 1.0) \times 10^{-7} \;\text{photon} \;\text{m}^{-2}\; \text{s}^{-1}$. 

\begin{figure*}[ht!]
\hspace{-0.5cm}
\fig{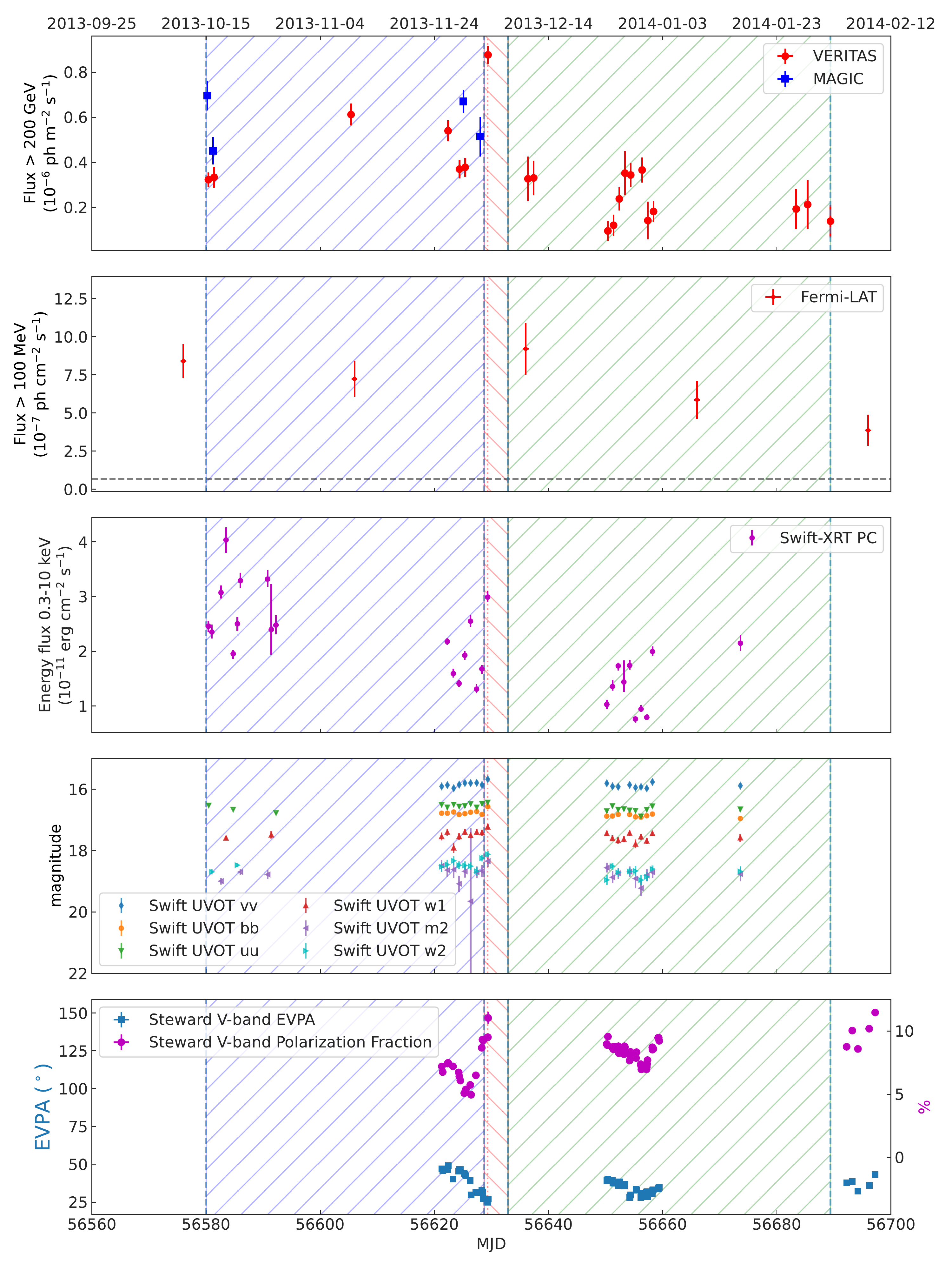}{0.9\textwidth}{}
\caption{MWL light curves of VER~J0521+211 focusing on the data between late 2013 and early 2014. The vertical dashed lines and the hatch fills illustrate the three Bayesian blocks (hatched filled in blue, red, and green, sequentially). The date ranges of the Bayesian blocks are listed in Table~\ref{tab:bb}. The vertical red dotted line illustrates the one night with the highest TeV gamma-ray flux, during which activity at other wavelengths can also be seen. Only 1-$\sigma$ statistical uncertainties are shown.  
\label{fig:lczoom}}
\end{figure*}

The TeV gamma-ray light curve ($>200$GeV) of VER~J0521+211 measured by VERITAS and MAGIC is shown in the top panel of Figure~\ref{fig:lc}. 
The TeV gamma-ray flux of the source was variable between 2013 Oct and 2014 Feb (see the top panel of Figure~\ref{fig:lczoom}), as a best-fit constant-flux model gives a $\chi^2$ value of 442.4 with 19 degrees of freedom (DOF), resulting in a negligible $p$-value. However, a fit of the TeV gamma-ray flux on the five nights with VERITAS observations between 2012 Nov and 2013 Feb to a constant-flux model yields a $\chi^2$ value of 9.1 with 4 DOF, equivalent to a $p$-value of 0.59. Therefore we do not rule out constant flux from the source in late 2012 through early 2013. 

The TeV gamma-ray flux above 200~GeV reached a peak of $(8.8 \pm 0.4) \times 10^{-7} \;\text{photon} \;\text{m}^{-2}\; \text{s}^{-1}$ ($\sim$37\% of the Crab Nebula flux), as measured by VERITAS on the night of 2013 Dec 3. No intra-day variability was found in the VERITAS data obtained on the night of the peak flux. 
On 2013 Dec 2, the night immediately before the peak TeV flux, MAGIC measured a lower flux above 200 GeV of $(5.1 \pm 0.9) \times 10^{-7} \;\text{photon} \;\text{m}^{-2}\; \text{s}^{-1}$.

The monthly-binned GeV gamma-ray light curve measured by {\it Fermi}-LAT between 2012 Oct and 2014 May is shown in the second panel from top of Figures~\ref{fig:lc} and \ref{fig:lczoom}. 
A fit of the GeV gamma-ray light curve to a constant-flux model yields a $\chi^2$ value of 102.5 with DOF$=22$, equivalent to a negligible $p$-value, showing that the source was variable during the period of interest. 
The average flux value during the one-year period between 2013 Jan 29 and 2014 Jan 24 was about 11 times that from the 3FGL catalog (illustrated by the horizontal dashed lines in Figures~\ref{fig:lc} and \ref{fig:lczoom}), and the monthly flux stayed above seven times the 3FGL catalog value, indicating a long-lasting elevated GeV gamma-ray flux state. 
During the period between 2013 Oct and 2014 Feb, when most of the observations at TeV gamma-ray and X-ray energies were performed, the GeV gamma-ray flux of the source did not show significant variability on monthly timescales. 

%%%%%%%%%
\subsubsection{Bayesian Blocks from the Gamma-Ray Light Curves} 
\label{subsec:GRBB}
%%%%%%%%%

A Bayesian blocks analysis \citep{Scargle13} of the TeV gamma-ray light curve was performed using the Python package Astropy \citep{Astropy13,Astropy2018} with an empirical prior equivalent to a false alarm probability $p_0=3\times10^{-7}$. Two change points (i.e., the edges of the Bayesian blocks), 2013 Dec 2 and 2013 Dec 6, were obtained from the VERITAS and MAGIC flux measurements, 
as illustrated by the light blue vertical dashed lines in Figure~\ref{fig:lc} and Figure~\ref{fig:lczoom}. These two change points define three Bayesian blocks, designated as BB1, BB2, and BB3, within each of which the flux is consistent with being constant and no variability is detected above 5-$\sigma$ statistical significance. The date ranges and the average fluxes of the three Bayesian blocks are shown in Table~\ref{tab:bb}. Note that there was only one night with 2.3 h of VERITAS observations on 2013 Dec 3 in the BB2 interval, with an averaged flux above 200~GeV of $(8.8 \pm 0.4) \times 10^{-7} \;\text{photon} \;\text{m}^{-2}\; \text{s}^{-1}$ ($\sim$37\% of the Crab Nebula flux), the highest of all three Bayesian blocks. 
The flux in BB3 is comparable to that in 2012, $(2.4 \pm 0.2) \times 10^{-7} \;\text{photon} \;\text{m}^{-2}\; \text{s}^{-1}$. 
MWL SEDs were constructed for the three time intervals from the Bayesian blocks analysis of the TeV gamma-ray light curve (see Section~\ref{sec:specvar}). 

\begin{table}[]
\centering
   \caption{ Bayesian blocks calculated from the TeV gamma-ray light curve shown in Figure~\ref{fig:lc}.  \\
   }
    \label{tab:bb}
\begin{tabular}{cccc}
\hline
 \hline
Period & Start & End & Flux ($>$200 GeV) \\
 & MJD & MJD & $10^{-7} \;\text{photon} \;\text{m}^{-2}\; \text{s}^{-1}$ \\ \hline
BB1 & 56580.0 & 56628.5 & $4.1 \pm 0.1$  \\
BB2 & 56628.5 & 56632.5 & $8.8 \pm 0.4$  \\
BB3 & 56632.5 & 56689.0 & $2.7 \pm 0.2$         \\ \hline
\end{tabular}
\end{table}

The same Bayesian block analysis was performed for the monthly binned \textit{Fermi}-LAT light curve. Two change points were found in the GeV flux of the source, 2013 Jan 29 and 2014 Jan 24. 
The GeV gamma-ray flux remained roughly constant in the year 2013, before and after which the flux tapered off. 

%%%%%%%%%
\subsubsection{X-ray and UV/optical Light Curves and Fractional Variability}
\label{subsec:fvar}

The light curves measured by the \textit{Swift}-XRT, the \textit{Swift}-UVOT, and the Steward Observatory are shown in the lower panels in Figures~\ref{fig:lc} and \ref{fig:lczoom}. The flux of the source at all of the studied wavelengths, as well as the optical polarization, both the electric vector position angle (EVPA) and the polarization fraction, exhibited variability during the period of interest.

\begin{figure}[ht!]
\hspace{-0.5cm}
\fig{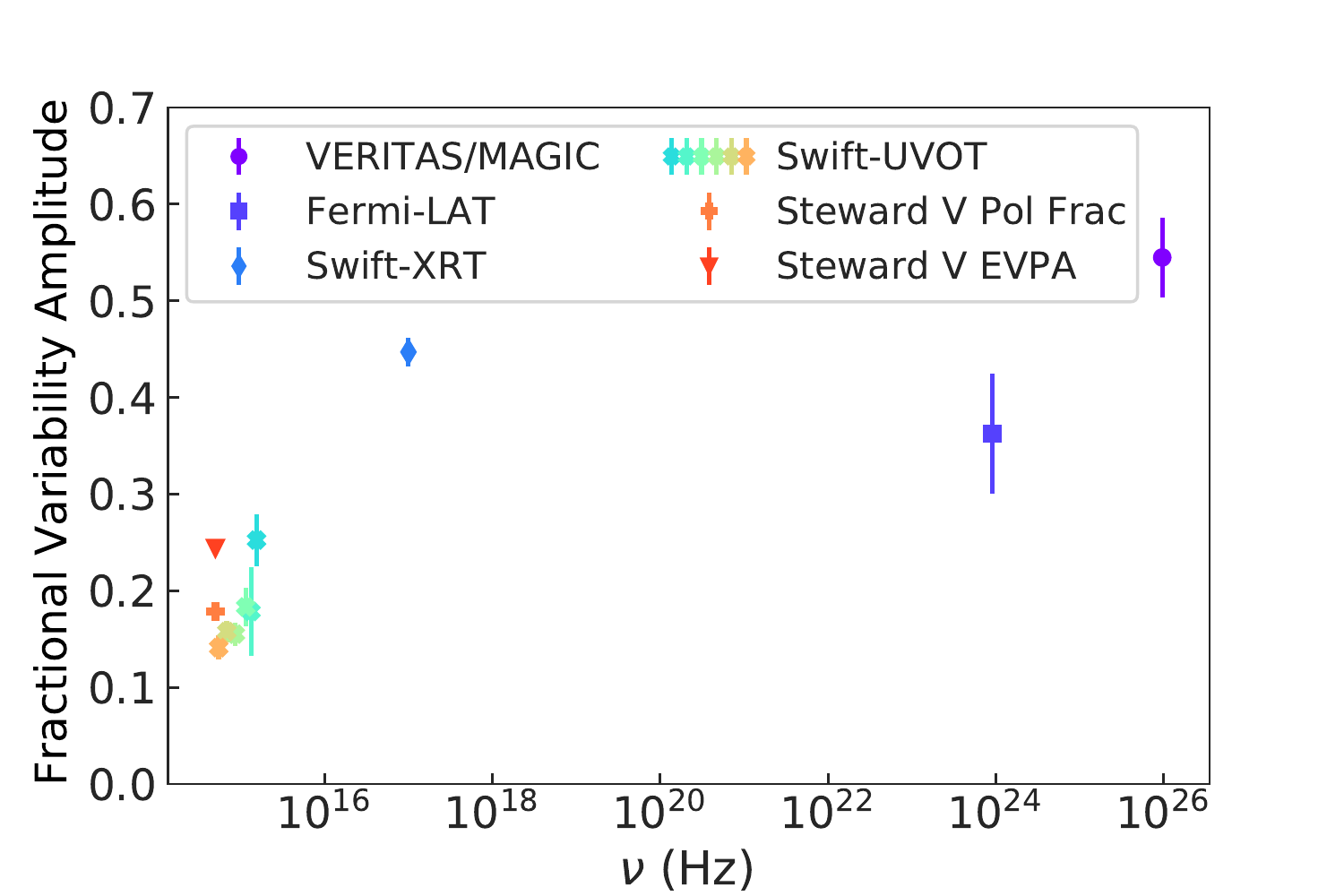}{0.5\textwidth}{}
\caption{
Fractional variability amplitude ($F_\text{var}$) as a function of frequency computed from the light curves shown in Figure~\ref{fig:lc}. 
The filled circle is calculated from the nightly binned VHE light curve, the filled square is from the monthly binned {\it Fermi}-LAT light curve, the filled diamond is from the {\it Swift}-XRT light curve binned by observations, the filled crosses are from the {\it Swift}-UVOT light curves with the six filters binned by observations, and the filled plus sign and downward triangle are from the $V$-band polarization fraction and EVPA measured by the Steward Observatory, respectively. 
}
\label{fig:fvar}
\end{figure}

To examine the relative variability amplitude at different wavelengths, the fractional variability amplitude $F_\text{var}$ \citep{Vaughan03, Poutanen08} as a function of frequency is shown in Figure~\ref{fig:fvar}. % and Table~\ref{tab:fvar}. 
The measured $F_\text{var}$ value calculated from each light curve is shown in Table~\ref{tab:fvar}. 
The VHE flux exhibits the highest fractional variability amplitude of $F_\text{var} = 0.54\pm0.03$ among all of the observed frequencies, with the X-ray flux yielding the second highest $F_\text{var} = 0.45\pm0.02$. The $F_\text{var}$ values in the optical and UV bands are the lowest. 

It is worth noting that $F_\text{var}$ depends on the timescales, both long (duration) and short (bin width), of the light curve. A wider sampling coverage of the same light curve with a smaller bin width and a longer duration will generally increase $F_\text{var}$. The durations of the light curves used to calculate $F_\text{var}$ are similar; so are the bin widths, with the exception of the monthly binned \textit{Fermi}-LAT light curve. 
\iffalse
Following the descriptions in \citet{Vaughan03} and \citet{Poutanen08}, the fractional variability $F_\text{var}$ and its error $\sigma_{F_\text{var}}$ are calculated as 
\[
F_\text{var}= \sqrt{\frac{S^2-\langle{\sigma_{err}^2}\rangle}{\langle F \rangle^2}}
\]
and 
\[
\sigma_{F_\text{var}}= \sqrt{F_\text{var}^2+\sqrt{\frac{2 \langle \sigma_{err}^2\rangle^2}{N\langle F\rangle^4} + \frac{ 4 \langle{\sigma_{err}^2}\rangle F_\text{var}^2}{N \langle F \rangle^2}}}-F_\text{var},
\]
where $S$ is the standard deviation of the N flux measurements, $\langle{\sigma_{err}^2}\rangle$ is the mean squared error of these flux measurements, $\langle F \rangle$ is the mean flux. 
\fi

\begin{table}[]
\centering
   \caption{ Fractional variability calculated from the MWL light curves shown in Figure~\ref{fig:lc}.  \\
   }
    \label{tab:fvar}
\begin{tabular}{lc}
\hline
 \hline
Instrument & $F_\text{var}$ \\ \hline
VERITAS and MAGIC & $0.54\pm0.03$  \\
\textit{Fermi}-LAT& $0.36\pm0.06$ \\
\textit{Swift}-XRT (PC) & $0.38\pm0.02$ \\
\textit{Swift}-XRT (WT) & $0.34\pm0.02$ \\
\textit{Swift}-XRT (combined) & $0.45\pm0.02$ \\
\textit{Swift}-UVOT ($UVW2$) & $0.25\pm0.03$ \\
\textit{Swift}-UVOT ($UVM2$) & $0.18\pm0.05$ \\
\textit{Swift}-UVOT ($UVW1$) & $0.18\pm0.02$ \\
\textit{Swift}-UVOT ($U$) & $0.16\pm0.01$ \\
\textit{Swift}-UVOT ($B$) & $0.16\pm0.01$ \\
\textit{Swift}-UVOT ($V$) & $0.14\pm0.01$  \\
Steward Observatory (Pol. Frac.) & $0.178\pm0.001$ \\
Steward Observatory (EVPA) & $0.244\pm0.001$         \\ \hline
\end{tabular}
\end{table}

%%%%%%%%%
\subsubsection{Optical Polarization Variability}
\label{subsec:optpolvar}
%%%%%%%%%

Significant variability in both the electric vector position angle (EVPA) and the polarization fraction were observed. 
The $V$-band EVPA roughly ranges between 25$^\circ$ and 50$^\circ$ during the time TeV gamma-ray observations were taken. The jet position angle from radio observations was measured in late 2012 and early 2014 to range from around 287$^\circ$ for the innermost feature at $\sim$5 mas (from the core) down to about 240$^\circ$ for the outermost feature at $\sim$50 mas \citep{Lister19}.
The polarization fraction varied between around 4\% and almost 12\%. On 2013 Dec 3 (the night with the highest TeV gamma-ray flux measured with VERITAS), a small increase in polarization fraction from $(9.47\pm0.08)\%$ to $(11.03\pm0.45)\%$ was observed. 
The fractional variability amplitudes $F_\text{var}$ of the $V$-band polarization fraction and EVPA are $0.178\pm0.001$ and $0.244\pm0.001$, respectively. 
Changes in polarization fraction were also observed about a month later, with possibly associated variability in X-ray and optical/UV fluxes around MJD 56660, but without significant flux elevation in the VHE band. 
Therefore, it is inconclusive whether or not the EVPA rotation and polarization fraction changes are associated with the gamma-ray variability.

It is worth noting that the behavior of the optical polarization of this source around the time of high TeV gamma-ray flux is quite different from the dramatic swing of the EVPA during some major gamma-ray flares observed for high-power blazars \citep[see e.g.,][]{Abdo10PolRot}. In those cases, a fast swing of the EVPA by 90$^\circ$ or 180$^\circ$ accompanied by a decrease (almost to 0) of the polarization fraction was observed during a major GeV gamma-ray flare. However, the relation between EVPA rotations, polarization fractions and gammma-ray flares in BL Lac objects appears to differ from case to case \citep[see e.g.,][]{Arlen13, MAGIC19BLLac}. 
%\citep{Zhang14}

%%%%%%%%%
\subsection{Spectral Variability}
\label{sec:specvar}
%%%%%%%%
\subsubsection{VHE Gamma-Ray Spectrum}
\label{subsec:GRSpec}
%%%%%%%%%

\begin{table*}[]
\centering
   \caption{ VHE gamma-ray spectral fit parameters using power-law and log-parabola models.  \\
   }
    \label{tab:VHEspecfit}
\begin{tabular}{cccccc}
\hline
 \hline
             & \multicolumn{2}{c}{Power Law} & \multicolumn{3}{c}{Log Parabola}         \\ %\hline
             & $\Gamma$      & $\chi^2$/DOF  & $\alpha$    & $\beta$     & $\chi^2$/DOF \\ \hline
VERITAS BB1  & $3.2\pm0.1$   & 4.3          & $3.2\pm0.1$ & $1.0\pm0.3$ & 0.8         \\
VERITAS BB2  & $3.1\pm0.1$   & 3.2          & $3.1\pm0.1$ & $0.9\pm0.2$ & 0.5         \\
VERITAS BB3  & $3.4\pm0.1$   & 1.6          & $3.1\pm0.1$ & $0.5\pm0.5$ & 1.7         \\
MAGIC Oct 15 & $3.0\pm0.1$   & 5.9           & $3.7\pm0.2$ & $1.9\pm0.4$ & 0.4          \\
MAGIC Oct 16 & $2.9\pm0.2$   & 0.8           & $3.1\pm0.3$ & $0.8\pm0.7$ & 0.6          \\
MAGIC Nov 29 & $3.1\pm0.1$   & 8.5           & $3.7\pm0.2$ & $1.7\pm0.3$ & 1.1          \\ \hline
\end{tabular}
\end{table*}

\begin{figure}[ht!]
\hspace{-1cm}
\gridline{\fig{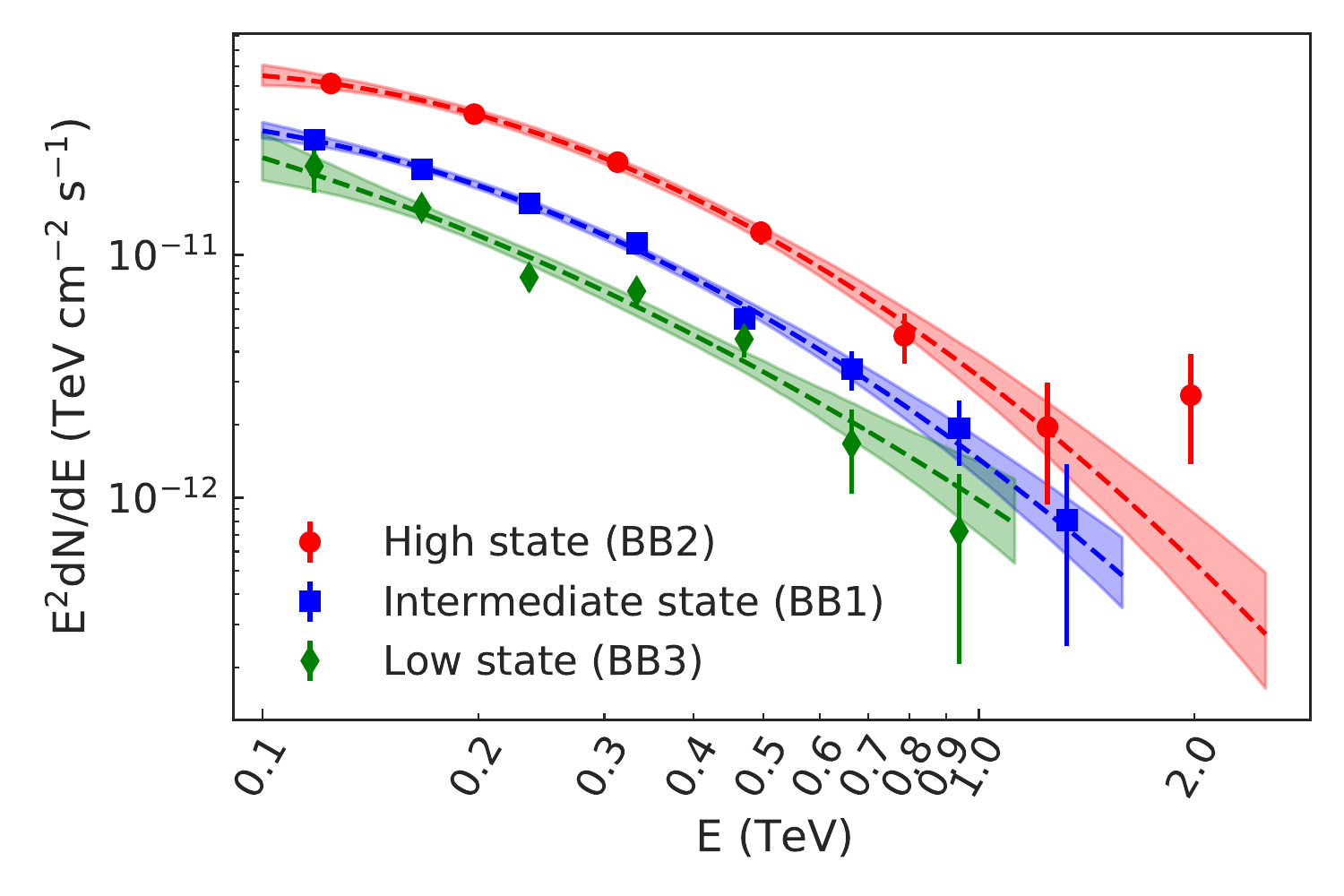}{0.45\textwidth}{}}
\caption{The VERITAS gamma-ray spectra of VER~J0521+211 during three different flux states. 
The dashed lines show the best-fit log-parabola models for the spectra.
The 1-$\sigma$  uncertainties from the model parameters are shown in shaded colors. These flux states were chosen based on a Bayesian block analysis of the VERITAS TeV gamma-ray light curve. 
\label{fig:vspec}}
\end{figure}

The VERITAS spectra during the three periods BB1 -- BB3, as defined in Section 3.1.2, are shown in Figure~\ref{fig:vspec}. 
No evidence of spectral variability was observed between these flux states, despite the integrated flux being different by a factor of four. 
The best-fit spectral parameters for a power-law model and a log-parabola model ($N_0(E/300~\text{GeV})^{[-\alpha-\beta \log_{10}(E/300~\text{GeV})]}$) for the three Bayesian blocks are shown in Table~\ref{tab:VHEspecfit}. 
The curved log-parabola model describes the spectra during BB1 and BB2 better than a power-law model. During the highest flux, in BB2, the VERITAS spectrum of the source extended beyond 1~TeV without any evidence of a sharp cutoff. 

%%%%%%%%%
%\subsection{MAGIC}
%%%%%%%%%

\begin{figure}[ht!]
\hspace{-0.5cm}
\fig{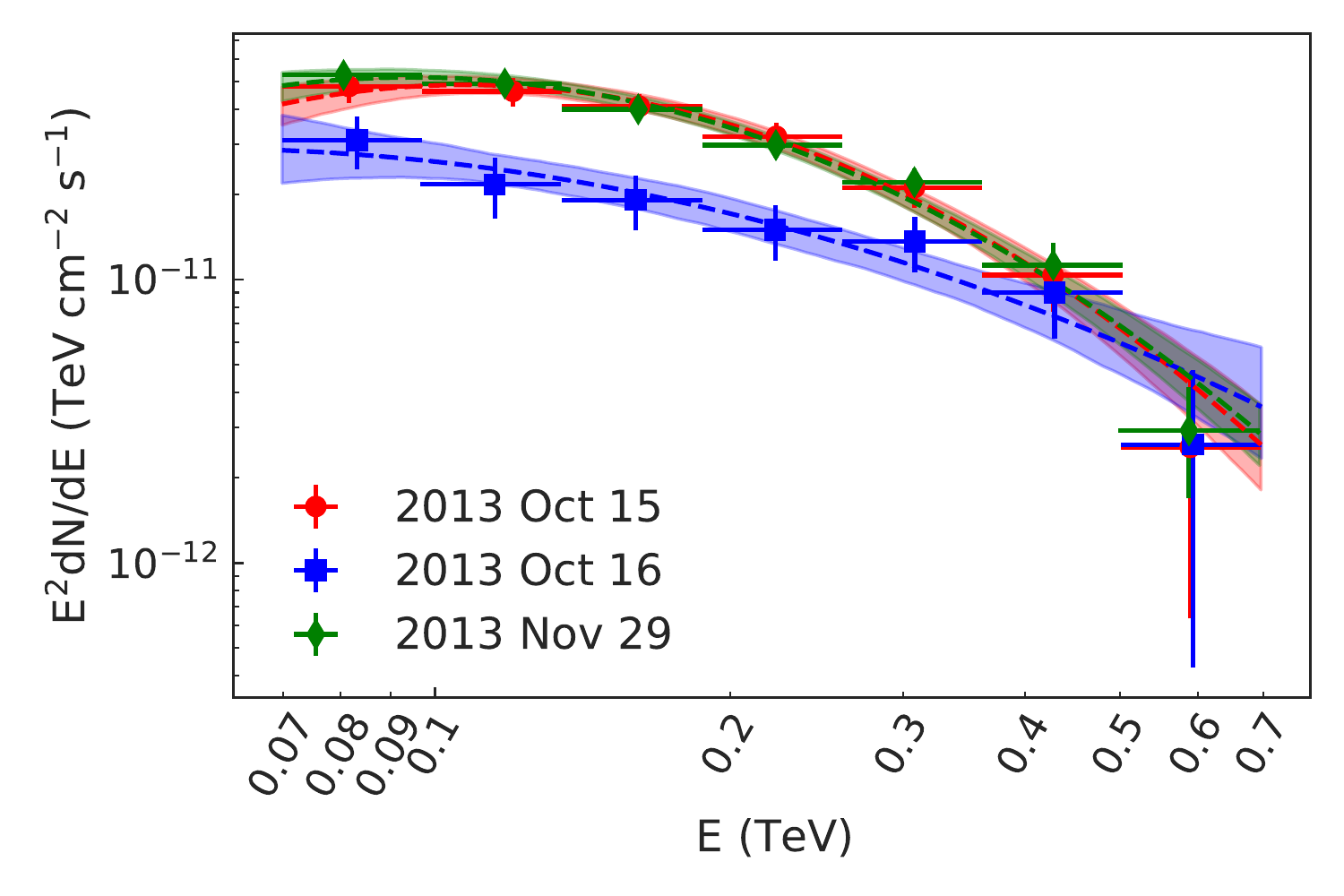}{0.45\textwidth}{}
\caption{The MAGIC gamma-ray spectra of VER~J0521+211 from observations on three nights in 2013. The dashed lines show the best-fit log-parabola models.
\label{fig:mspec}}
\end{figure}

The MAGIC spectra measured on three of the four nights, 2013 Oct 15, Oct 16, and Nov 29, are shown in Figure~\ref{fig:mspec}, together with a log-parabola fit. 
The best-fit spectral parameters and reduced $\chi^2$ values for these three nights are shown in Table~\ref{tab:VHEspecfit}. 
Note that the MAGIC spectra extend down to a lower energy (70 GeV) than those from VERITAS, thanks to the larger areas of the reflectors, capturing slightly more spectral curvature at low energies.

%%%%%%%%
\subsubsection{X-Ray and UV/Optical Spectral Variability}
\label{subsec:XRTvar}

\begin{figure}[ht!]
\hspace{-0.3cm}
\fig{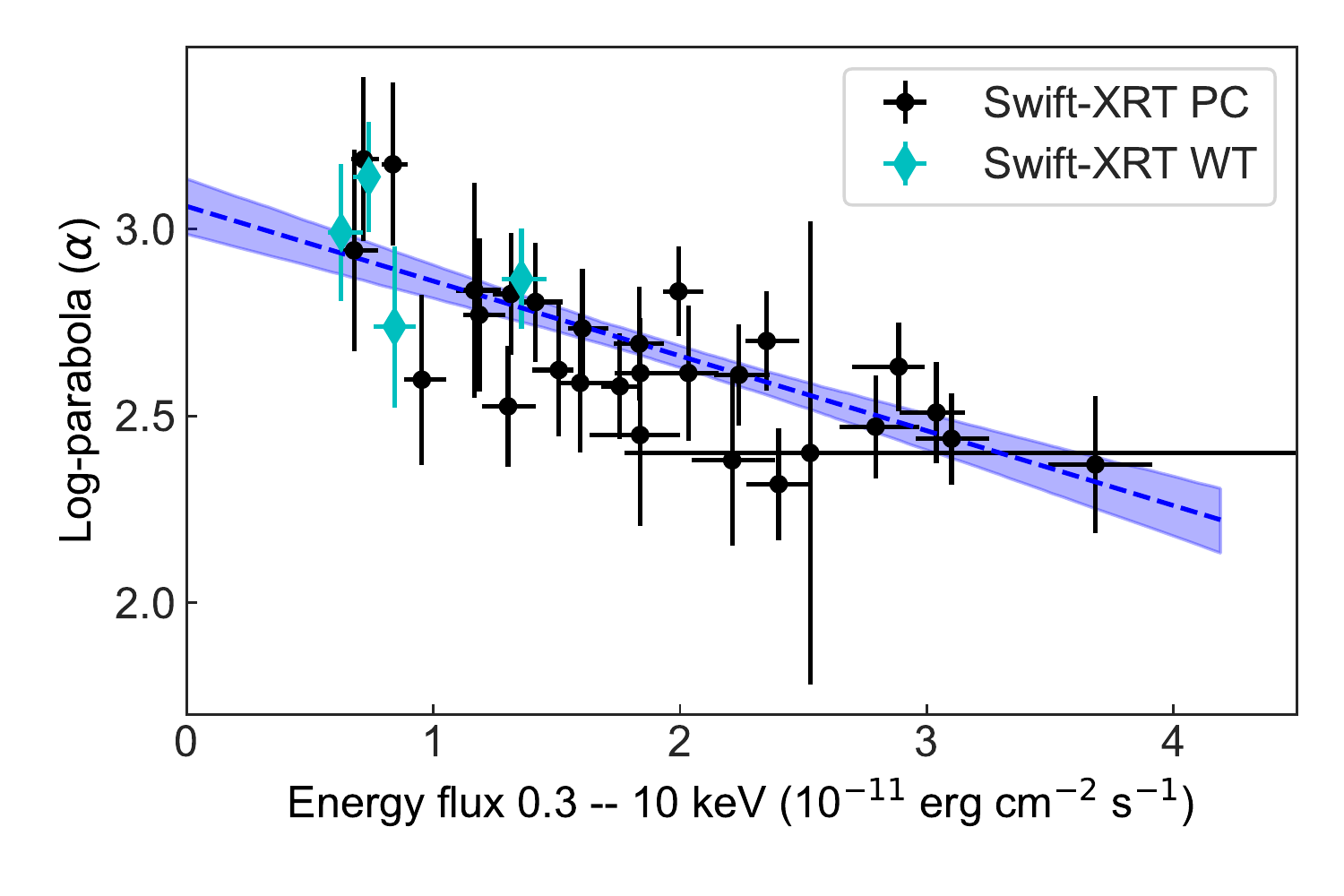}{0.51\textwidth}{}
\hspace{-0.2cm}
\caption{Scatter plot to investigate the correlation between the best-fit X-ray photon index at 2 keV and the X-ray energy flux. The blue dashed line shows the best-fit linear model.  
\label{fig:xrt_flux_index}}
\end{figure}

The X-ray spectrum from each {\it Swift}-XRT observation was extracted. The detailed information of these observations and 
the results of the spectral fits are shown in Table~\ref{tab:Xspec} in the Appendix. 
The relation between the best-fit photon index at 2 keV ($\alpha+\beta\log2$) and the energy flux in XRT data is illustrated by Figure~\ref{fig:xrt_flux_index}. The log-parabola energy-dependent photon index at 2 keV is comparable to the power-law index for this source, and therefore chosen to investigate the X-ray index-flux correlation. A ``harder-when-brighter'' trend, i.e., a smaller photon index when the flux is higher, is apparent. The Pearson correlation coefficient between the index and flux is 0.76 ($p$-value $\sim6\times10^{-7}$). A linear fit yields a slope of $-0.2\pm0.04$. 

A combined analysis was performed using observations taken during each of the three Bayesian blocks. 
The best-fit power-law model suggests that the X-ray spectrum during the last Bayesian block was marginally softer (with an index at 2 keV of $2.73\pm0.11$) than that for the first Bayesian block (with an index at 2 keV of $2.57\pm0.06$). 

The UV/optical spectrum of the source can be obtained by combining the extinction-corrected fluxes at the six wavelengths from the {\it Swift}-UVOT filtered observations. 
However, the uncertainty in the measured reddening affects the extinction correction, most notably in the UV band, and subsequently affects the interpretation of the frequency of the synchrotron peak in the broadband SED. 
If the true reddening is closer to the value $E(B-V)$=0.703 from \citet{SFD1998}, the frequency of the synchrotron peak of the SED would be above the UV band, making this source an HBL during the flaring state. However, if the true reddening is the value $E(B-V)$=0.605 from \citet{SandF2011}, as we used in this work, the frequency of the synchrotron peak would be below $10^{15}$ Hz, still in the IBL regime. 

Because of the uncertainty in the measured reddening, we did not include data taken with the two {\it Swift}-UVOT filters with the highest frequencies in the UV/optical spectra. 
The UVOT spectra at the lower four frequencies during the three periods were fitted with a power-law model to estimate the spectral variability. The UV spectrum is marginally harder during BB2, with a best-fit index of $2.23\pm0.09$, whereas those during BB1 and BB3 are $2.41\pm0.06$ and $2.37\pm0.05$, respectively. This is consistent with the ``harder-when-brighter'' behavior in the X-ray band. Because the extinction correction is the same for all periods, the relative change is robust against the chosen reddening value. 

%%%%%%%%
\subsubsection{X-Ray/Gamma-Ray Correlation}
\label{subsec:XVcorr}

\begin{figure}[ht!]
\hspace{-0.5cm}
\fig{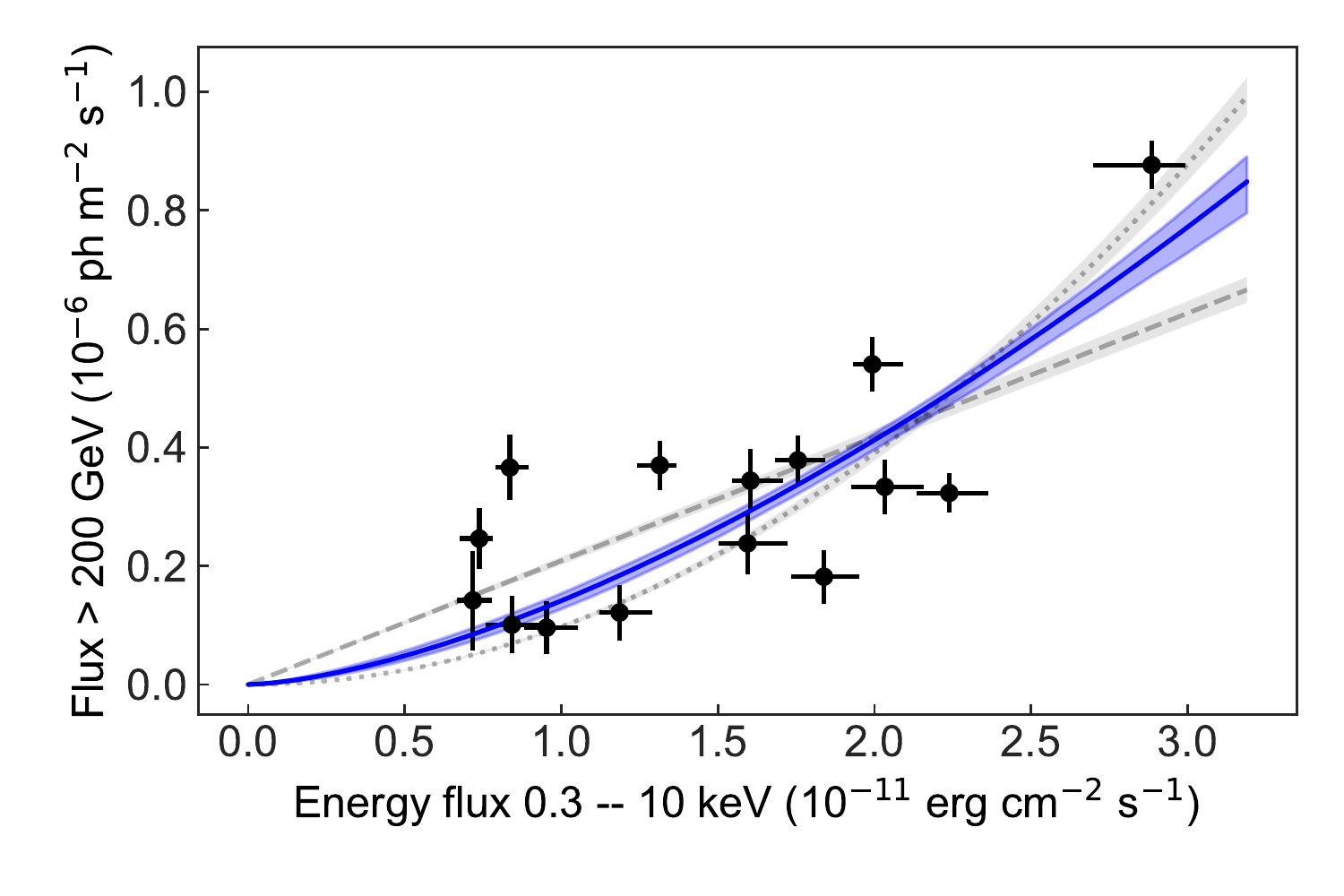}{0.51\textwidth}{}
\caption{Scatter plot to investigate the correlation between the TeV gamma-ray flux and the X-ray energy flux. The flux measurements in the two bands were performed on the same nights. The gray dashed and dotted lines show the best linear and quadratic fits, respectively. The solid blue line shows the best power-law fit. 
\label{fig:xvcorr}}
\end{figure}

A moderate correlation between TeV gamma-ray and X-ray fluxes observed within a one-day window is apparent, as shown in Figure~\ref{fig:xvcorr}, with a Pearson correlation coefficient $r$ between the two bands of 0.74 ($p$-value 0.0014). Note that the small number of data points and the potential non-Gaussian distribution of the observed fluxes of the source makes it hard to interpret the Pearson $r$ correlation. 
A linear fit, a quadratic fit, and a power-law fit ($F_\text{TeV}\propto F_\text{X}^a$) were performed. All three models yielded rather large reduced $\chi^2$ values (between 8.4 and 9.6), likely due to the intrinsic scatter of the gamma-ray/X-ray correlation. The best-fit power-law index is $a=1.55 \pm 0.12$, indicating the correlation between TeV gamma-ray and X-ray fluxes is likely between quadratic ($a=2$) and linear ($a=1$). 

%%%%%%%%%
\subsubsection{Optical--UV/Gamma-Ray Correlation}
\label{subsec:OptUVGRcorr}
%%%%%%%%%

\begin{figure}[ht!]
\hspace{-0.5cm}
\fig{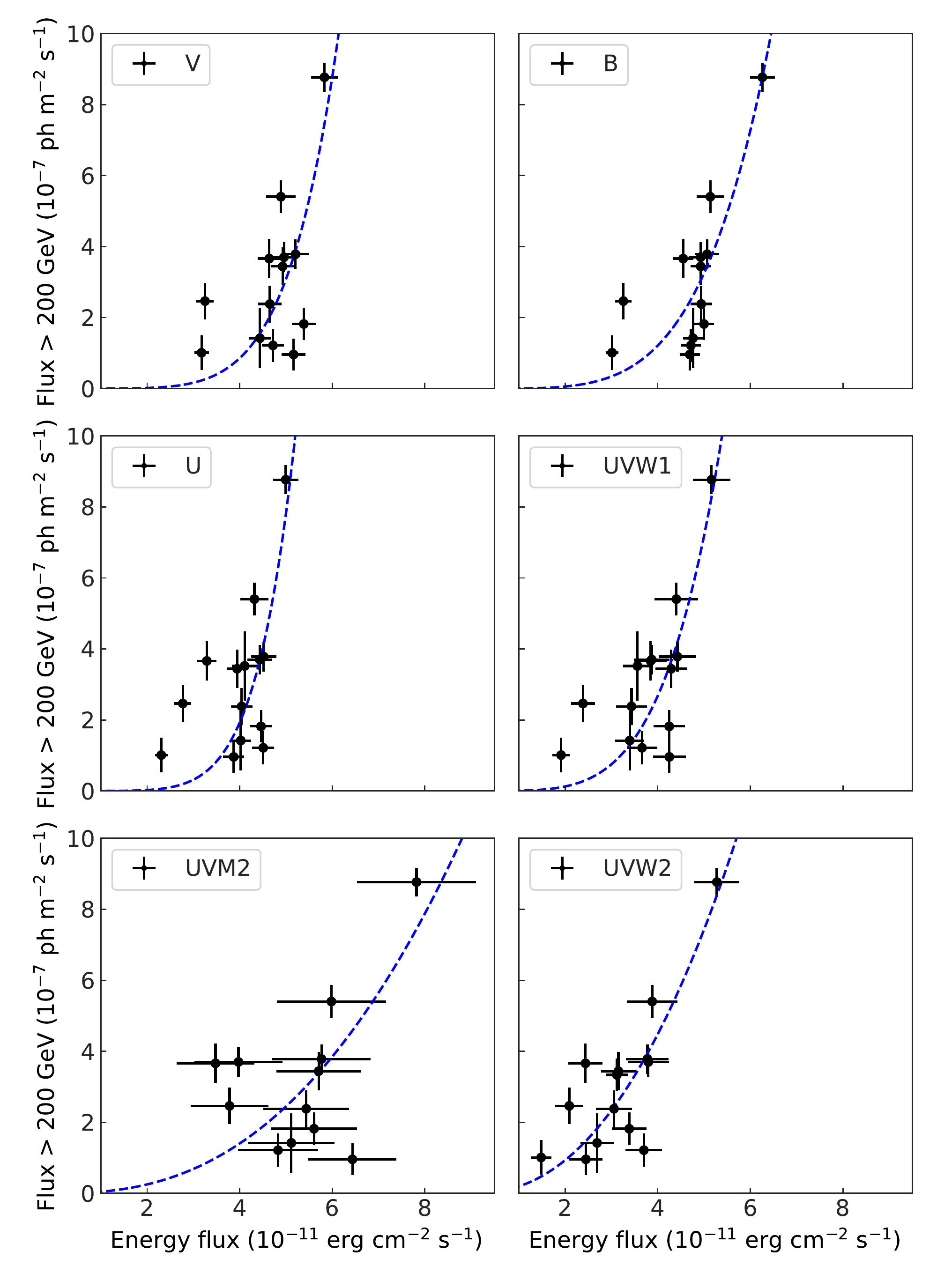}{0.5\textwidth}{}
\caption{
The correlations between the TeV gamma-ray flux and the optical/UV energy fluxes measured with the six {\it Swift}-UVOT filters. The frequency of the {\it Swift}-UVOT observations increases from the upper left panel to the lower right panel. Similar to Figure~\ref{fig:xvcorr}, the flux measurements in the two bands in each panel were performed on the same nights. The blue dashed lines show the best-fit power law to guide the eye. 
\label{fig:uvcorr}}
\end{figure}

The correlations between the TeV gamma-ray flux and the optical/UV energy fluxes measured with the six {\it Swift}-UVOT filters are shown in Figure~\ref{fig:uvcorr}.
The Pearson correlation coefficients range between 0.47 and 0.76 (the $p$-values range between 0.01 and 0.08 except for the $UVW2$ filter), reaching the highest value of 0.76 (a $p$-value of 0.0015) between the UV flux observed with the highest-frequency $UVW2$ filter and the TeV gamma rays (as shown in the lower right panel of Figure~\ref{fig:uvcorr}). A power-law fit ($F_\text{TeV}\propto F_\text{UVOT}^b$) indicates a more-than-quadratic relation with indices $b>2$ between these bands, although the goodness-of-fit is poor with reduced $\chi^2$ values ranging between 6.4 (between $UVW2$ and VHE) and 16.7 (between $UVM2$ and VHE). However, the fractional variability amplitudes $F_\text{var}$ in the optical/UV bands are much lower than in the VHE and X-ray bands, increasing with frequency from $0.14\pm0.01$ for the lowest-frequency $V$ filter to $0.25\pm0.03$ for the highest-frequency $UVW2$ filter, as shown in Figure~\ref{fig:fvar}. 
The low $F_\text{var}$ values for optical/UV fluxes would generally lead to a steeper optical/UV/VHE correlation with a large index $b$. 
We note that the $F_\text{var}$ values are not affected by extinction correction. 

%%%%%%%%%
\subsection{Broadband SED}
\label{subsec:SED}
%%%%%%%%%

\begin{figure*}[ht!]
\hspace{-1cm}
\fig{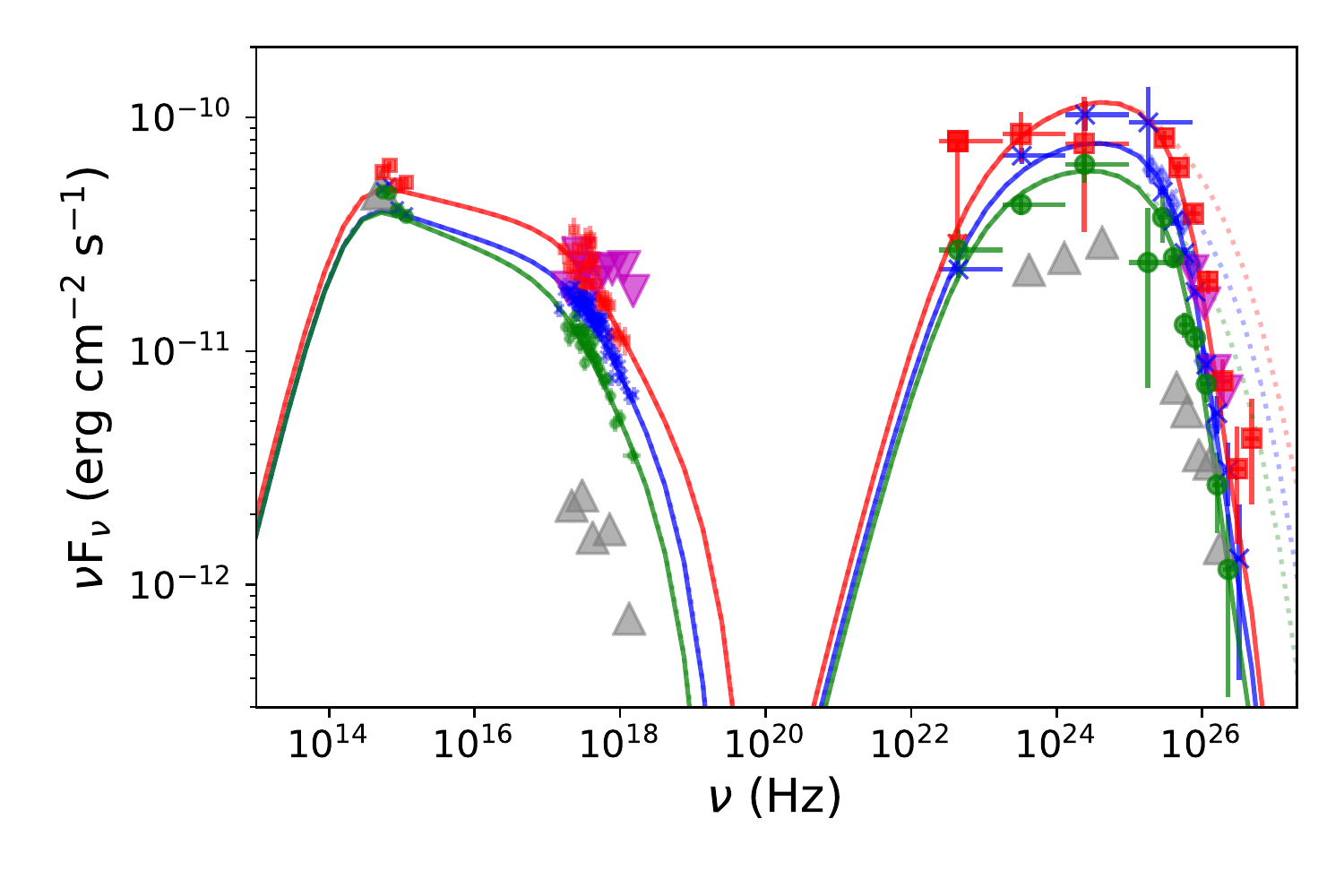}{1.0\textwidth}{}
\caption{Broadband SED during the three Bayesian blocks obtained from the VERITAS light curve in the 2013 to 2014 season. The solid and dotted curves show the one-zone SSC model with the parameters listed in Table~\ref{tab:SSCmodel} with and without the extragalactic background light (EBL) absorption \citep{Franceschini08}. 
The blue crosses and filled diamonds, the red squares and the green filled circles represent the first (BB1), second (BB2, flaring), and the third (BB3) Bayesian blocks, respectively. Data are from {\it Swift}-UVOT, {\it Swift}-XRT, {\it Fermi}-LAT, and MAGIC (only for BB1, shown as blue filled diamonds) and VERITAS, going from the low to high frequencies. 
As a comparison, the magenta downward triangles and gray upward triangles show the flaring-state and low-state SEDs in 2009 from \citet{Archambault13}. 
\label{fig:sed}}
\end{figure*}

The broadband SEDs were constructed for the three Bayesian block intervals, BB1, BB2, and BB3, and are shown in Figure~\ref{fig:sed}. The EBL absorption of the VHE gamma-ray spectra were considered for following \citet{Franceschini08}, assuming a redshift of $z=0.18$ \citep{Paiano17}. The X-ray spectra were corrected for the neutral hydrogen absorption (see Section~\ref{subsec:XRT}). The UV/optical spectra were corrected for extinction (see Section~\ref{subsec:UVOT}). The SEDs are modeled with a one-zone SSC model as described below in Section~\ref{subsec:seddis}. 

%%%%%%%%%%%%%%%%%%
%%%%%%%%%%%%%%%%%%
%%%% %     Discussion     %%%%
%%%%%%%%%%%%%%%%%%
%%%%%%%%%%%%%%%%%%
\section{Discussion}

%%%%%%%%%
\subsection{Flux Variability}
\label{subsec:vardis}
%%%%%%%%%

The blazar VER~J0521+211 was observed at an elevated gamma-ray state between 2013 and 2014 with VERITAS, MAGIC, and {\it Fermi}-LAT. 
The TeV gamma-ray flux above 200~GeV reached a peak of $(8.8 \pm 0.4) \times 10^{-7} \;\text{photon} \;\text{m}^{-2}\; \text{s}^{-1}$ ($\sim$37\% of the Crab Nebula flux) on the night of 2013 Dec 3. 
The MAGIC observations on 2013 Dec 2, one day before the night of the peak TeV gamma-ray flux, yielded a flux above 200 GeV of $(5.1 \pm 0.9) \times 10^{-7} \;\text{photon} \;\text{m}^{-2}\; \text{s}^{-1}$, implying variability on a daily timescale. 
The GeV gamma-ray flux above 100 MeV stayed above five times the value from 3FGL for over five months between 2013 Sep and 2014 Feb, with an average flux of 10 times the 3FGL value. 

MWL variability from the optical to VHE bands was observed in 2013 and 2014, with the fractional variability amplitude $F_\text{var}$ increasing with frequency from the lowest $0.14\pm0.01$ for $V$-band flux to $0.45\pm0.02$ for X-ray flux, and then decreasing to $0.36\pm0.06$ for GeV gamma-ray flux, before increasing again to the highest value $0.54\pm0.03$ for TeV gamma-ray flux. Such a pattern in $F_\text{var}$ is consistent with previous observations of gamma-ray blazars \citep[see e.g., ][and references therein]{Balokovic16}, with the highest $F_\text{var}$ observed at the frequencies above the two peaks in the SED.

%%%%%%%%%
\subsection{Cross-band Correlations}
\label{subsec:corrdis}
%%%%%%%%%
%
A moderate correlation between the TeV gamma-ray and X-ray fluxes from the source was observed. 
The best-fit power law model suggests that the correlation is more than linear, with an index of $1.47\pm0.13$, consistent with the expectation of a varying electron density and adiabatic cooling in a SSC model \citep{Krawczynski02, Katarzynski05}.
The TeV gamma-ray and X-ray correlation could also give insights into whether the IC scattering occurs in the Thomson regime or KN regime, as the correlation may become closer to linear in the KN regime with the diminished scattering cross-section \citep[e.g.][]{Aleksic15M42010,Balokovic16}. 
More data are needed to quantify the correlation between X-rays and TeV gamma-rays with a higher precision. 

The correlation between the optical and TeV gamma-ray fluxes was weaker than that between X-ray and TeV gamma-ray fluxes, while the correlation between UV and TeV gamma-ray fluxes was comparable to that between the X-rays and TeV gamma rays, with a slightly higher correlation coefficient observed between the UVOT measurements with the UVW2 filter at the highest central frequency of $\sim1.55\times 10^{15}$~Hz. 
The UV/TeV gamma-ray correlation is more than quadratic, which could occur if the emitting region expands as the particle density increases \citep{Katarzynski05}. 

%%%%%%%%%
\subsection{Upper limits on the Redshift}
\label{subsec:zdis}
%%%%%%%%%
Despite multiple measurement attempts \citep{Shaw13,Paiano17}, the redshift of VER~J0521+211 remains uncertain. 
To facilitate the discussion, 
we adopt the value $z=0.18$, the lower limit reported by \citet{Paiano17}. At this redshift, the luminosity distance is 878 Mpc (using $\Omega_m=0.29$, $\Omega_{\Lambda}=0.71$, and $H_0=69.6\;\text{km}\;\text{s}^{-1}\;\text{Mpc}^{-1}$ \citep{Bennett2014}), and the angular scale is roughly 3 pc mas$^{-1}$. 

To help constrain the redshift of the source, 
upper limits were derived from the gamma-ray spectra using two methods based on \citet{Mazin07} and \citet{Georganopoulos10}. 

The first method \citep{Mazin07} assumes that the intrinsic gamma-ray spectrum follows a power law with an index of $\Gamma_\text{int}>1.5$ (or more conservatively, $\Gamma_\text{int}>0.7$). This method was used to derive the 95\% upper limit on the redshift of $z<0.34$ (and $z<0.44$ for the conservative assumption of $\Gamma_\text{int}>0.7$) for VER~J0521+211 from VERITAS observations in 2009 and 2010 \citep{Archambault13}. Using the same method, we find that the VERITAS spectrum on the night of the flare (BB2) offers a stronger constraint than other time periods, yielding a 95\% upper limit on the redshift of $z<0.38$ (and $z<0.53$ for the conservative assumption of $\Gamma_\text{int}>0.7$). 
We note that the deabsorbed VERITAS spectra at $z=0.38$ are convex, with the measured fluxes at high energies significantly exceeding the best-fit power-law model, indicating that the redshift of the source is likely to be well below these limits. 

\begin{figure}[ht!]
\hspace{-0.5cm}
\gridline{\fig{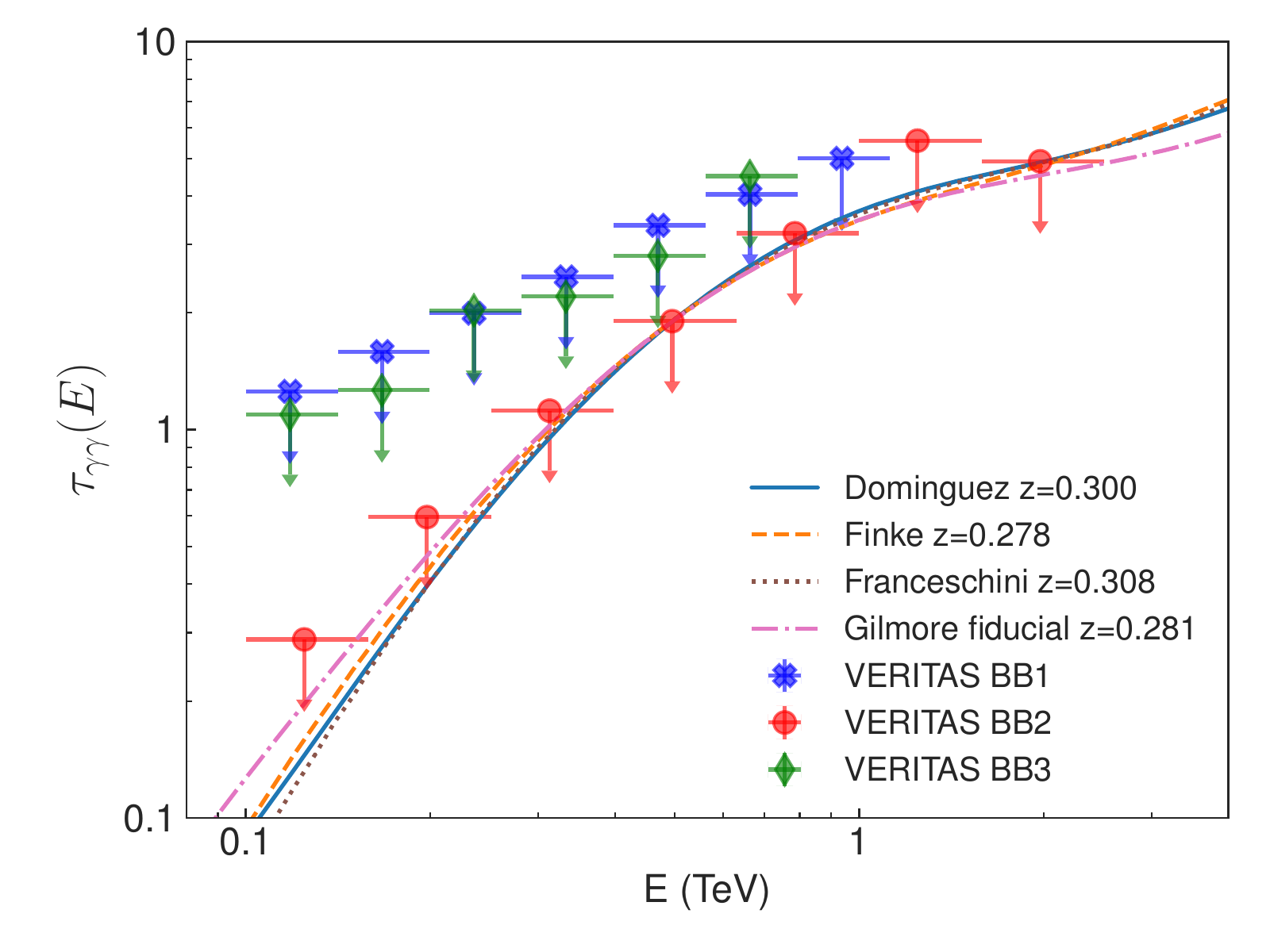}{0.45\textwidth}{}}
\gridline{\fig{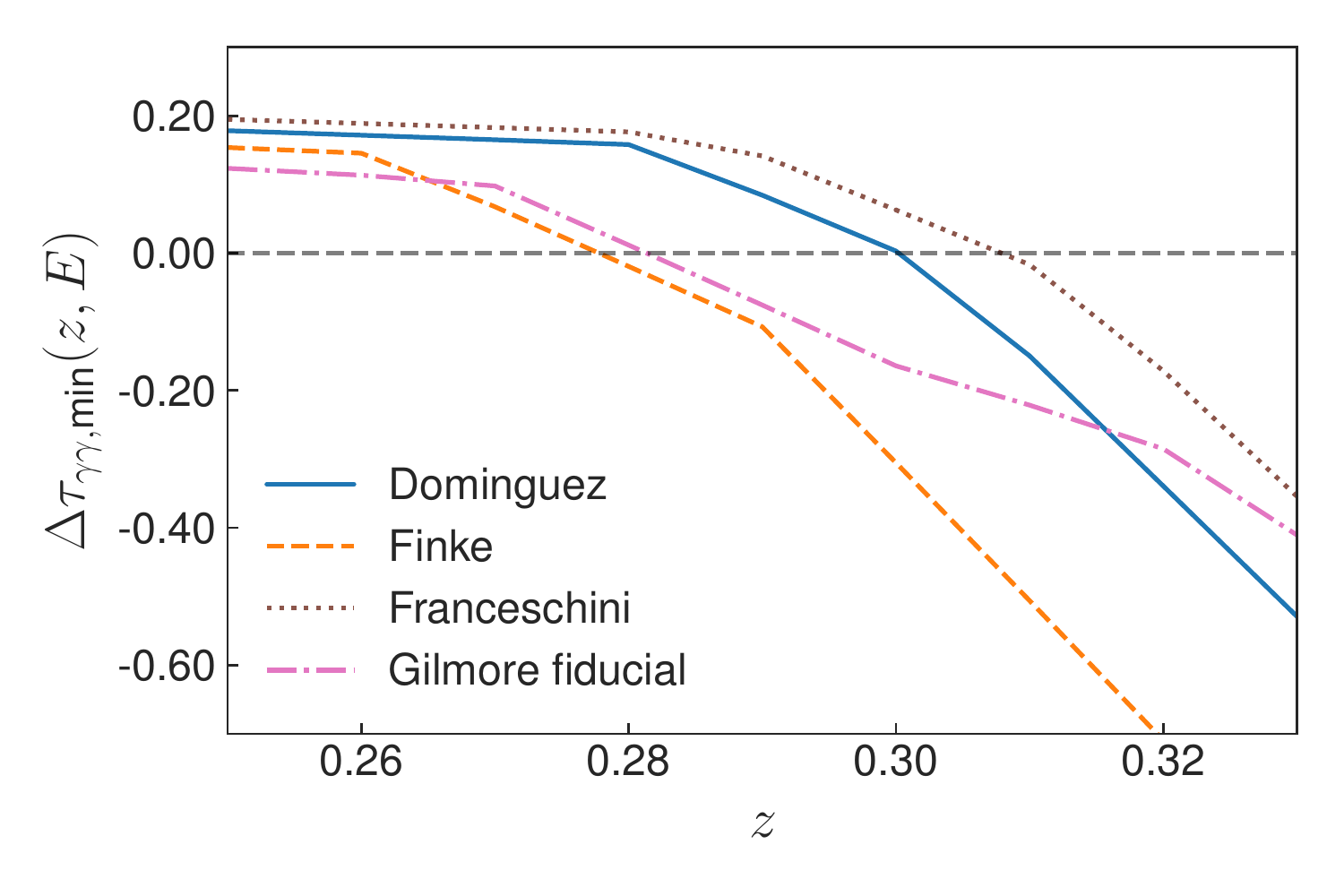}{0.45\textwidth}{}}
\caption{{\it(Upper)} The optical depth $\tau_{\gamma\gamma}(E)$ of the gamma rays due to $\gamma\gamma$ pair absorption induced by EBL photons. The blue filled crosses, the red filled circles, and the green filled diamonds show the 95\% upper limits derived from VERITAS observations during periods BB1, BB2, and BB3. The optical depth $\tau_{\gamma\gamma}(z, E)$ from four EBL models, \citet{Dominguez11} (blue solid line), \citet{Finke10} (orange dashed line), \citet{Franceschini08} (copper dotted line), and \citet{Gilmore12} (pink dash dotted line), evaluated at the maximum redshift allowed by the 95\% upper limits are also shown. 
{\it(Lower)} The minimum difference between the 95\% upper limits on the EBL-induced pair absorption optical depth $\tau_{\gamma\gamma}(E)$ calculated from VERITAS data and the optical depth $\tau_{\gamma\gamma}(z, E)$ calculated from EBL models as a function of redshift. }
\label{fig:redshift}
\end{figure}

The second method \citep{Georganopoulos10} assumes that the intrinsic TeV gamma-ray flux does not exceed the extrapolation of the GeV gamma-ray spectrum, which is minimally affected by the EBL for TeV blazars. 
A 95\% upper limit on the optical depth from the EBL absorption can be calculated following Equation~2 in \citet{Aleksic11} 
\begin{linenomath}
\[
\tau_{\gamma\gamma}(E) \le \log [\frac{F_\text{int}(E)}{F_\text{obs}-1.64\Delta F_\text{obs}}],
\]
\end{linenomath}
where $F_\text{int}(E)$ is the extrapolated GeV flux, representing the maximally allowed intrinsic flux, at energy $E$, $F_\text{obs}$ is the observed TeV gamma-ray flux, and $\Delta F_\text{obs}$ is the uncertainty of $F_\text{obs}$. The 95\% upper limits on the optical depth $\tau_{\gamma\gamma}(E)$ derived from {\it Fermi}-LAT and VERITAS spectra for periods BB1, BB2, and BB3 are shown in the upper panel of Figure~\ref{fig:redshift}. The spectrum from BB2 offers the strongest constraints on $\tau_{\gamma\gamma}(E)$. The optical depths $\tau_{\gamma\gamma}(z, E)$ for sources at a range of redshift of $0.25\le z \le 0.35$ from four EBL models \citep{Dominguez11,Finke10, Franceschini08,Gilmore12} were computed and compared to the constraints from VERITAS observations. The minimum difference between the observational constraints and the model-predicted optical depth, $\Delta \tau_{\gamma\gamma,\text{min}}(z, E)$ as a function of redshift is shown in the lower panel of Figure~\ref{fig:redshift}. The 95\% upper limits on the redshift of the source (corresponding to $\Delta \tau_{\gamma\gamma,\text{min}}(z, E) = 0$) were found to be $z\le0.3$, $z\le0.278$, $z\le0.308$, and $z\le0.281$ for the models from \citet{Dominguez11}, \citet{Finke10}, \citet{Franceschini08}, and \citet{Gilmore12}, respectively. These upper limits on the redshift are more constraining than the first method and those reported by \citet{Archambault13}. A precise measurement of the spectroscopic redshift of this source in the future will be important, as it helps constrain both the particle distribution and the EBL intensity. 

%%%%%%%%%
\subsection{An SSC Model for the Broadband SED}
\label{subsec:seddis}
%%%%%%%%%
Three periods were identified with a Bayesian block analysis during which the source exhibited different TeV gamma-ray flux states 
(BB1, BB2, and BB3 as mentioned above). 
Broadband SEDs from UV to TeV gamma rays were constructed for each period. 
These SEDs are adequately described by a one-zone SSC model calculated following \citet{Krawczynski02}. 
The sets of parameters used to generate the SEDs shown in Figure~\ref{fig:sed} are listed in Table~\ref{tab:SSCmodel}. 
These parameters differ from those in the SSC model for the low-state SED from VERITAS data in 2009 \citep{Archambault13}, where an even lower magnetic field strength of $B=0.0025$~G results in a long synchrotron cooling time and a particle-dominated jet away from equipartition. 

The simple SSC modeling provides several constraints on the properties of the source that are discussed below.

%Classification
The peak frequency of the synchrotron radiation from the SSC model is roughly $5\times 10^{14}$~Hz, in the optical band and the IBL regime. However, as mentioned in Section~\ref{subsec:UVOT}, the synchrotron peak frequency closely depends on the intrinsic optical/UV spectrum, and is heavily affected by extinction correction. If more substantial extinction from dust exists, the synchrotron peak would be in the UV band and the source in the HBL regime. 

We note that it has been observed that the synchrotron peak frequency can shift between different flux states for blazars \citep[e.g.,][]{Acciari11501, Valverde2020}. 
Particularly, VER~J0521+211 has been reported to exhibit HBL-like properties during a previous flare on 2009 November 27 \citep{Archambault13}. The ``flare'' and ``low-state'' SEDs from \citet{Archambault13} are shown, as a comparison to this work, as the magenta downward triangles and gray upward triangles in Figure~\ref{fig:sed}. 
The TeV gamma-ray spectrum, including normalization and index, during the entire period studied in this work was comparable to that during the ``flaring'' state in 2009 November. 
However, while the X-ray spectra in this work were similar in normalization, they were much softer than that during the ``flaring'' state in 2009 November (with a photon index value of $2.0\pm0.1$) \citep{Archambault13}. This suggests strong synchrotron X-ray spectral variability on timescales of years for VER~J0521+211. 
Moreover, the optical fluxes in this work were comparable to the low-state flux in \citet{Archambault13}, with a rather low variability amplitude. 
As the optical flux is only mildly affected by the extinction correction, this low variability amplitude in optical band is robust. 

The low optical variability implies that the SED at/below the synchrotron peak frequency does not change as much as that above the peak. 
The relatively flat spectrum from UV to X-ray bands indicates that the synchrotron peak is likely broad, making it difficult to measure the peak frequency and classify the blazar. 
On the other hand, the UV/X-ray spectrum of the source becomes harder with higher flux (as illustrated by the UV photon indices in Section~\ref{subsec:UVOT} and the index-flux relation in Figure~\ref{fig:xrt_flux_index}). 
The similar behavior of the observed UV and X-ray fluxes implies that the optical/UV bands are also likely located above the synchrotron SED peak, confirming the source as an IBL even in the higher flux state in 2013. %
The ``harder-when-brighter" trend in both bands could be caused by the synchrotron SED peak shifting towards higher frequencies during flares. 
Such a behavior of the SED can be explained by changes to particles with the highest energies, e.g., from an evolving maximum or break particle energy or a change in the particle distribution shape above the break. 

As the synchrotron peak frequency of VER J0521+211 is around the borderline between IBL and HBL, we review the radio features in the innermost jet region, which also provide insights into the blazar subclasses. 
\citet{Lister19} identified six moving features in the inner jet from 15 GHz VLBA observations since 2009, including one that moves at a large apparent speed of $\sim 0.77\pm0.05$ mas yr$^{-1}$ at $\sim$5 mas from the core in mid 2014. On the other hand, three innermost features move at a very small apparent velocity, two of which even exhibited statistically significant inward motion. These slow features may be quasi-stationary.  %(the angular size scale at $z=$0.108 is roughly 2 pc mas$^{-1}$). 
Such a combination of quasi-stationary and fast moving radio knots is typically found in IBL/LBLs \citep[e.g.][]{Hervet16}. 
Therefore, the MOJAVE observations are consistent with the source being an IBL.

Interestingly, a 15-GHz moving feature at an apparent speed of $\sim 0.21\pm0.02$ mas yr$^{-1}$ first appeared at around 5.5 mas ($\sim$11 pc) in 2012 December \citep{Lister2018},\footnote{\url{http://www.physics.purdue.edu/astro/MOJAVE/sepvstime/0518+211_sepvstime.png}} coincident with the period when the source exhibited elevated gamma-ray flux. Similar coincidences have been observed before in IBLs, but between fast gamma-ray flares on intra-day timescales and moving knots at distances much closer to the core observed with the VLBA at 43 GHz \citep[e.g., from BL Lacertae][]{Arlen13, Abeysekara18}. 
The interactions between the moving and stationary radio features in the jet could provide a possible particle acceleration mechanism that leads to enhanced gamma-ray emission.

%\begin{rotatetable}
%\begin{landscape}
\begin{table*}
\centering
   \caption{ Parameters used for the one-zone SSC models shown in Figure~\ref{fig:sed}.  \\
   {\footnotesize $\Gamma$ and $\theta$ are the bulk Lorentz factor and the angle with respect to the line-of-sight of the relativistic emitting region, respectively; $R$ is the radius of this region; $B$ is the strength of the magnetic field; $w_e$ is the energy density of the emitting electrons; $\log E_\text{min}$, $\log E_\text{max}$, and $\log E_\text{break}$ are the logarithms of the minimum, maximum, and break electron energies, respectively; $p_1$ and $p_2$ are the spectral indices of the electrons below and above the break energy, respectively. The corresponding parameters listed in Table~3 in \citet{Archambault13} for the low-state SED in 2009 are also included for a comparison. }
   }
    \label{tab:SSCmodel}
\begin{tabular}{cccccccccccc} \\ \hline \hline
State & $\Gamma$ & $\theta$ & $B$ & $R$ & $w_e$ & \multirow{2}{*}{$\log \frac{E_\text{min}}{\text{eV}} $} & \multirow{2}{*}{$\log \frac{E_\text{max}}{\text{eV}}$} & \multirow{2}{*}{$\log \frac{ E_\text{break}}{\text{eV}}$} & $p_1$ & $p_2$  \\
      &         & deg   &  $10^{-2}$ G & $10^{17} $ cm      & $10^{-3}$ erg cm$^{-3}$ &    &        &          &    & \\ \hline
% below are v44 z=0.18 (R_17)
  BB1   & 25 & 2.2 & 1.5 & 1.01 & 1.32 & 9.7 & 12.0 & 11.25 & 3.2 & 4.2  \\
  BB2   & 25 & 2.2 & 1.5 & 1.05 & 1.45 & 9.7 & 12.2 & 11.25 & 3.15 & 4.15 \\
  BB3   & 25 & 2.2 & 1.5 & 1.1 & 1.0 & 9.7 & 11.9 & 11.15 & 3.25 & 4.25 \\ %\hline
  2009$^1$ & 30 & - & 0.25 & 4.0 & - & 10.25 & 12.0 & - & 3.0 & - \\ \hline
\end{tabular}\\
  \begin{minipage}{0.95\textwidth}%
    \footnotesize $^1$ \citet{Archambault13}. %
  \end{minipage}%
\end{table*}

Based on the SSC model parameters (see~Table~\ref{tab:SSCmodel}), we can investigate the emission process of the particles at the source. 
The Lorentz factor of those electrons emitting the peak synchrotron radiation 
$\gamma_\text{syn} \approx 5.4 \times 10^4 (B/1\;G)^{-1/2} [\delta/(1+z)]^{-1/2} (\nu_\text{syn}/10^{16}\;\text{Hz})^{1/2} \approx 2 \times 10^{4}$ (for BB1/BB3), where $B$ is the strength of the magnetic field, $\delta$ is the Doppler factor ($\delta = [\Gamma (1-\beta cos \theta)]^{-1}=26$, where $\beta = (1-1/\Gamma^2)^{1/2}$), $z$ is the redshift, and $\nu_\text{syn}$ is the peak frequency of the synchrotron radiation. 
The synchrotron cooling time of these electrons emitting at the peak frequency is $t_\text{syn}\approx 7.74\times 10^8\;\text{s}\times (B/1\;\text{G})^{-2} \times \gamma^{-1} \approx 2000 \;\text{days}$, corresponding to $\sim$78 days in the observer's frame. The electrons responsible for the peak synchrotron emission during BB2 have a slightly higher Lorentz factor of $2.1 \times 10^{4}$ and a cooling time of 73 days in the observer's frame. 
As the synchrotron radiation energy density in the model is a few times larger than the magnetic field energy density, the cooling is likely dominated by the inverse-Compton scattering, and the total observed cooling time can be as short as two weeks. 
Since the dynamic timescale is approximately $R/c\sim$1.5~day in the observers' frame, the adiabatic cooling could also dominate the decay of the flux of the source. 

Using the Lorentz factor of the electrons emitting at the synchrotron peak frequency $\gamma_\text{syn}$, we checked for the importance of the KN effects, which become substantial when the photon energy $h\nu^\star>m_e c^2$ in the electron rest frame. 
Considering the IC scattering between the photons at the synchrotron peak frequency in the comoving frame $E'_\text{peak}\approx 0.1$~eV (primed quantities refer to those in the comoving frame) and the electrons that produce the peak synchrotron photons, we have $\gamma_\text{syn}E'_\text{peak}\approx2$~keV $<m_e c^2$. Therefore, the IC scattering of the peak of the synchrotron radiation field by electrons with $\gamma_\text{syn}$ occurs in the Thomson regime \citep{Acciari11501} and produces the observed IC radiation up to $4\delta\gamma_\text{syn}^2E'_\text{peak}\approx4$~GeV, below the high-energy SED peak. 

From the SED modeling, the SSC component peaks at a frequency $\nu_\text{ssc}\approx 4.4\times10^{24}$~Hz for the first two blocks, and moves to a lower frequency $\nu_\text{ssc}\approx 2.6\times10^{24}$~Hz for the last, low-flux block. The shift of the SSC peak to higher frequencies during flares has been previously observed in blazars \citep[e.g.,][]{Acciari11501}. 
We also calculate the limit of the KN Doppler factor \citep[Eq. (17) in][]{Tavecchio98} $\delta < \delta_\text{KN}=[4\nu_\text{syn}\nu_\text{ssc}/(3(m_ec^2/h)^2)]^{1/2} \allowbreak \exp{\{1/(\alpha_1-1)  +1/[2(\alpha_2-\alpha_1)]\}}^{-1/2}$, where the spectral indices below and above the IC peak are taken from {\it Fermi}-LAT and VERITAS observations as $\alpha_1\approx0.9$ and $\alpha_2\approx2.1$, respectively. 
In this case, the Doppler factor $\delta=26$ is below the KN limit $\delta_\text{KN}$ ranging between about 40 and 60. As a result, the inverse Compton scattering that produces the peak SSC emission occurs in the KN regime. 
Note, however, that the observed SSC peak lies around 15~GeV, therefore the best-fit index from the {\it Fermi}-LAT spectrum may not be a good estimator of $\alpha_1$. Moreover, the KN limit Doppler factor, defined as Eq. (17) in \citet{Tavecchio98}, depends sensitively on $\alpha_1$. If we use a value of $\alpha_1\approx0.8$ as an example, the KN limit Doppler factor becomes about 6, placing the gamma-ray SED peak in the Thomson regime. 

The maximum electron energy $E_{e, \text{max}}$ ranges between $10^{11.9}\;\text{eV}$ and $10^{12.2}\;\text{eV}$ ($\gamma$ between about $1.6\times10^6$ and $3.1\times10^6$), 
and the observed maximum gamma-ray energy is $E_\gamma \sim 1$ TeV, corresponding to an energy of  $E_\gamma' = E_\gamma (1+z) / \delta \sim 40$~GeV in the comoving frame. Therefore, we can estimate the energy range of the low-energy photons that participate in inverse-Compton scattering and produce the observed gamma rays to be roughly between $E_\gamma' m_e^2c^4 / E_{e, \text{max}}^2 \approx 0.01$~eV and $m_e^2c^4/E_\gamma' \sim 6.8$~eV in the comoving frame. These correspond to the observed photon energies between 0.25 and 175~eV, or approximately $6 \times 10^{13}$ and $4 \times 10^{16}$~Hz, ranging from the IR/optical bands up to the UV and soft X-ray band. In this scenario, we expect the observed emission at some of these lower energies to be strongly correlated with the gamma-ray emission. If the target photons of the observed energies above $\delta m_e c^2 / \gamma_\text{max} \sim$4~eV were up-scattered by the electrons at the maximum energy, the inverse-Compton scattering would occur in the KN regime, leading to spectral breaks in the synchrotron spectrum that are not observed. 
However, the observed correlations between optical/UV and VHE fluxes (as shown in Figure~\ref{fig:uvcorr}) are not stronger than the observed X-ray/VHE correlation (as shown in Figure~\ref{fig:xvcorr}), implying that the lower-energy electrons and the target photons with observed energies above $\sim$6.4~eV (corresponding to the frequency of the $UVW2$ filter) are probably responsible for the observed VHE flux. 

The electron break energy $E_\text{brk}$ ranges between $10^{11.15}$~eV and $10^{11.25}$~eV ($\gamma$ between $2.8\times10^5$ and $3.5\times10^5$). These electrons emit synchrotron radiation at around $2\times10^{17}$~Hz (roughly 1~keV), and have a synchrotron cooling time of a few days in the observers' frame. 
Assuming the mean free path ($\lambda$) of these electrons is comparable to their gyroradius of between $3.1 \times 10^{10}$ cm and $3.9 \times 10^{10}$ cm, we can estimate the electron diffusion coefficient $D \approx \lambda^2 / (2\lambda/c)$ to be a few times $10^{20} \;\text{cm}^2 \;\text{s}^{-1}$. Taking the radius of the emitting region used in the SED modeling, we can estimate the escape timescale of the electrons at the break energy $\tau_\text{esc} \approx R^2/(4D)$ to be on the order of $10^5$ yr in the comoving frame, much longer than the cooling time.   

The light crossing time in the observer's frame is $2R'(1+z)/(c\delta)\sim 3$ days, shorter than the shortest Bayesian block interval ($\sim$5 days), and the model obeys causality. 

A leptonic model has been used to describe the observed SED from VER J0521+211, allowing information about the jet environment to be inferred.  It is possible that protons are accelerated in this same jet environment.
By requiring protons to be magnetically confined within the emitting region \citep{Hillas84}, a rough estimation of the maximum proton energy can be calculated as 
$E_{p}<3\times10^{-11}\;\text{PeV}(B/1\;\text{G})(R/1\;\text{m})\approx500\;\text{PeV}$. If photohadronic processes occur for protons at the maximum energy, the corresponding maximum energy of the produced neutrinos is then $E_\nu\le\gamma_p m_\pi c^2/4\approx 0.04E_p\approx20\;\text{PeV}$ \citep{Rachen98}, where $\gamma_p$ is the Lorentz factor of the protons and $m_\pi$ is the mass of a charged pion. 
Since no considerations were made of the acceleration, radiation, or escape processes of the protons, the maximum proton energy could be lower due to the constraints from these processes. However, we do not rule out the possibility of neutrino production at energies up to 20~PeV in VER~J0521+211 from the confinement argument alone. 
Given the relatively long duration of the elevated TeV gamma-ray flux, VER~J0521+211 is potentially a good candidate for astrophysical neutrino searches. 

%%%%%%%%%%%%%%%%%%
%%%%%%%%%%%%%%%%%%
%%%% %     Summary     %%%%
%%%%%%%%%%%%%%%%%%
%%%%%%%%%%%%%%%%%%
\section{Summary}
The blazar VER~J0521+211 was observed at an elevated gamma-ray state between 2013 and 2014 with VERITAS, MAGIC, and {\it Fermi}-LAT. The TeV gamma-ray flux above 200~GeV reached a peak of $(8.8 \pm 0.4) \times 10^{-7} \;\text{photon} \;\text{m}^{-2}\; \text{s}^{-1}$ ($\sim$37\% of the Crab Nebula flux), and the monthly GeV gamma-ray flux above 100 MeV remained higher than seven times the 3FGL catalog value for a period that lasted one year. 

The fractional variability amplitude $F_\text{var}$ was observed to be the lowest for the lowest-frequency $V$-band flux, and increases with frequency to the X-ray band, and then decreases with frequency to the GeV energies, before increasing again to the highest at the highest TeV energies. Such a pattern of high variability amplitude $F_\text{var}$ being observed at frequencies above the two peaks in the SED is consistent with previous observations of gamma-ray blazars \citep[see e.g., ][and references therein]{Balokovic16}. 

A moderate more-than-linear correlation between the X-ray and TeV gamma-ray fluxes was observed, with a Pearson correlation coefficient of 0.7. The X-ray spectrum appeared harder when the flux was higher. Unlike the gamma-ray and X-ray band, the optical flux did not increase significantly during the studied period compared to the archival low-state flux. 
The combination of low optical variability amplitude, the higher X-ray variability amplitude, and the ``harder-when-brighter'' trend in the X-ray spectrum indicates that the synchrotron peak of the SED may become broader during flaring states of this source, and the synchrotron peak frequency can shift to higher frequencies. 
Such a behavior of the SED can be explained by activity from particles with the highest energies. 
Of particular interest, the X-ray spectrum in 2013 and 2014 is much softer than that observed in a flaring state in 2009 \citep{Archambault13}, while the normalization values are comparable, suggesting strong X-ray spectral variability on timesales of years. 

The correlation between the optical and VHE fluxes is weaker than that between X-ray and VHE fluxes, while that between UV and VHE fluxes is the strongest and follows a more-than-quadratic relation, possibly from the combination of a varying particle density and an expansion of the emitting region \citep{Katarzynski05}. 
The weaker correlation between the optical and VHE fluxes, as well as the low optical variability observed from 2009 to 2014, could also suggest that the optical emission from the source originates from a different region \citep[see e.g., ][]{Rajput19}.

Upper limits at the 95\% confidence level on the redshift of the source were derived following \citet{Georganopoulos10} and \citet{Aleksic11} to be $z\le0.3$, $z\le0.278$, $z\le0.308$, and $z\le0.281$ using four EBL models from \citet{Dominguez11}, \citet{Finke10}, \citet{Franceschini08}, and \citet{Gilmore12}, respectively. These upper limits are more constraining than the previously reported value of $z<0.34$ \citep{Archambault13}. A precise redshift measurement of this source in the future will be important to studies of the particle distribution as well as the EBL intensity. 

The broadband SED can be adequately described with a one-zone synchrotron self-Compton model. The inverse-Compton scattering in the source happens in the Thomson regime below the high-energy SED peak and likely transitions to the KN regime at or above the high-energy SED peak.
In contrast to the hard X-ray spectrum and HBL-like SED properties of VER J0521+211 observed during the flaring state in 2009, the soft X-ray spectrum, the SED synchrotron peak frequency, the ``harder-when-brighter" trend in both UV and X-ray band, and the radio morphology from the MOJAVE program are all consistent with the source being an IBL object during the flaring state in 2013. 

As a rare IBL that can be detected at TeV gamma-ray energies on timescales of months, more simultaneous MWL observations of VER~J0521+211 can provide valuable insights into this type of source. Of particular interest are the gamma-ray band and X-ray band, as they have exhibited more variability compared to other energies. Continued radio interferometry monitoring is also interesting for IBLs, as there could be a potential association between the ejection of superluminal knots and gamma-ray flares. 

\begin{acknowledgments}
This research is supported by grants from the U.S. Department of Energy Office of Science, the U.S. National Science Foundation and the Smithsonian Institution, by NSERC in Canada, and by the Helmholtz Association in Germany. This research used resources provided by the Open Science Grid, which is supported by the National Science Foundation and the U.S. Department of Energy's Office of Science, and resources of the National Energy Research Scientific Computing Center (NERSC), a U.S. Department of Energy Office of Science User Facility operated under Contract No. DE-AC02-05CH11231. We acknowledge the excellent work of the technical support staff at the Fred Lawrence Whipple Observatory and at the collaborating institutions in the construction and operation of the instrument. 
\end{acknowledgments}
\begin{acknowledgments}
The MAGIC Collaboration would like to thank the Instituto de Astrof\'{\i}sica de Canarias for the excellent working conditions at the Observatorio del Roque de los Muchachos in La Palma. The financial support of the German BMBF, MPG and HGF; the Italian INFN and INAF; the Swiss National Fund SNF; the ERDF under the Spanish Ministerio de Ciencia e Innovaci\'on (MICINN) (PID2019-104114RB-C31, PID2019-104114RB-C32, PID2019-104114RB-C33, PID2019-105510GB-C31,PID2019-107847RB-C41, PID2019-107847RB-C42, PID2019-107847RB-C44, PID2019-107988GB-C22); the Indian Department of Atomic Energy; the Japanese ICRR, the University of Tokyo, JSPS, and MEXT; the Bulgarian Ministry of Education and Science, National RI Roadmap Project DO1-400/18.12.2020 and the Academy of Finland grant nr. 320045 is gratefully acknowledged. This work was also supported by the Spanish Centro de Excelencia ``Severo Ochoa'' (SEV-2016-0588, SEV-2017-0709, CEX2019-000920-S), the Unidad de Excelencia ``Mar\'{\i}a de Maeztu'' (CEX2019-000918-M, MDM-2015-0509-18-2) and by the CERCA program of the Generalitat de Catalunya; by the Croatian Science Foundation (HrZZ) Project IP-2016-06-9782 and the University of Rijeka Project uniri-prirod-18-48; by the DFG Collaborative Research Centers SFB823/C4 and SFB876/C3; the Polish National Research Centre grant UMO-2016/22/M/ST9/00382; and by the Brazilian MCTIC, CNPq and FAPERJ.
\end{acknowledgments}
\begin{acknowledgments}
Qi Feng was supported by NSF Grant PHY-1806554 at Barnard College.
\end{acknowledgments}

%\vspace{5mm}
\facilities{VERITAS, MAGIC, Fermi(LAT), Swift(XRT), Steward Observatory, VLBA}

\software{Astropy \citep{Astropy13,Astropy2018},  
          NumPy \citep{numpy11},
          Matplotlib \citep{Hunter07},
          SciPy \citep{SciPy},
          seaborn \citep{seaborn14}, 
          HEASoft, VEGAS \citep{Cogan08}, Eventdisplay \citep{Maier17}, 
          Fermitools
          }

\appendix
%%%%%%%%%
%%%%%%%%%

The X-ray spectral fit results using a log-parabola model for each observation are shown in Table~\ref{tab:Xspec}. 
\begin{deluxetable*}{cccccchc}
\tablecaption{Results from X-ray spectral analysis ($N_H = 4.38 \times 10^{21}$ cm$^{-2}$) \label{tab:Xspec}}
\tabletypesize{\scriptsize}
\tablehead{
\colhead{Observation} 	&\colhead{Start time} 	&\colhead{Exposure} 	&\colhead{Energy flux}  						& \colhead{$\alpha$}   & \colhead{$\beta$}     		&\nocolhead{$N_H$} 					& $\chi^{2}/\text{DOF}$                            \\
\colhead{} 			&\colhead{}  			&\colhead{(ks)} 		&\colhead{($10^{-12}$ ergs cm$^{-2}$ s$^{-1}$)}  	& \colhead{}    	& \colhead{}    				&\nocolhead{($10^{21}$ cm$^{-2}$)} 		& \colhead{}               
}
\startdata
32628001 & 2012-11-13 09:46:51 &             2.0 &  $8.4^{+0.8}_{-0.9}$ & 3.00$\pm$0.22 & -0.37$\pm$0.40 & $4.4^{+4.4}_{--4.4}$ &  21.1/27 \\
32628002 & 2012-11-14 08:14:59 &             2.3 &  $7.4^{+0.6}_{-0.4}$ & 3.65$\pm$0.15 & -0.74$\pm$0.31 & $4.4^{+4.4}_{--4.4}$ &  56.4/35 \\
32628003 & 2012-11-15 08:17:59 &             1.8 &  $6.3^{+0.5}_{-0.9}$ & 3.83$\pm$0.18 & -1.21$\pm$0.38 & $4.4^{+4.4}_{--4.4}$ &  35.2/22 \\
32628004 & 2012-11-16 08:34:59 &             1.9 & $13.6^{+0.8}_{-1.0}$ & 3.19$\pm$0.14 & -0.46$\pm$0.27 & $4.4^{+4.4}_{--4.4}$ &  53.2/44 \\
31531010 & 2013-10-15 09:22:59 &             1.5 &  $22.4^{+1.0}_{-1.2}$ & 2.10$\pm$0.14 &  0.74$\pm$0.24 & $4.4^{+4.4}_{--4.4}$ &  40.9/42 \\
31531011 & 2013-10-15 23:58:59 &             1.3 &  $20.3^{+1.1}_{-1.2}$ & 1.87$\pm$0.18 &  1.08$\pm$0.33 & $4.4^{+4.4}_{--4.4}$ &  23.8/32 \\
31531012 & 2013-10-17 14:13:59 &             1.6 &  $27.9^{+1.5}_{-1.8}$ & 1.98$\pm$0.14 &  0.70$\pm$0.23 & $4.4^{+4.4}_{--4.4}$ &  55.1/44 \\
31531013 & 2013-10-18 11:29:59 &             0.7 &  $36.8^{+1.9}_{-2.3}$ & 1.80$\pm$0.18 &  0.82$\pm$0.29 & $4.4^{+4.4}_{--4.4}$ &  23.4/31 \\
31531014 & 2013-10-19 17:31:17 &             1.4 &  $18.3^{+1.0}_{-1.0}$ & 2.31$\pm$0.15 &  0.56$\pm$0.27 & $4.4^{+4.4}_{--4.4}$ &  29.9/33 \\
31531015 & 2013-10-20 11:05:59 &             1.6 &  $24.0^{+1.3}_{-1.3}$ & 2.13$\pm$0.15 &  0.26$\pm$0.25 & $4.4^{+4.4}_{--4.4}$ &  42.1/38 \\
31531016 & 2013-10-21 00:01:58 &             1.4 &  $31.0^{+1.4}_{-1.5}$ & 2.11$\pm$0.12 &  0.47$\pm$0.20 & $4.4^{+4.4}_{--4.4}$ &  66.4/53 \\
31531017 & 2013-10-25 18:04:59 &             1.3 &  $30.4^{+1.5}_{-1.1}$ & 2.06$\pm$0.14 &  0.64$\pm$0.24 & $4.4^{+4.4}_{--4.4}$ &  63.5/48 \\
31531018 & 2013-10-26 10:06:29 &             0.1 & $25.3^{+7.5}_{-45.3}$ & 2.51$\pm$0.62 & -0.16$\pm$1.75 & $4.4^{+4.4}_{--4.4}$ &    0.0/1 \\
31531019 & 2013-10-27 05:14:59 &             1.4 &  $22.1^{+1.6}_{-1.8}$ & 1.87$\pm$0.23 &  0.74$\pm$0.39 & $4.4^{+4.4}_{--4.4}$ &  17.4/21 \\
31531020 & 2013-11-25 05:28:59 &             2.1 &  $11.7^{+0.7}_{-1.1}$ & 1.63$\pm$0.29 &  1.73$\pm$0.51 & $4.4^{+4.4}_{--4.4}$ &  35.6/18 \\
31531021 & 2013-11-26 05:30:59 &             2.1 &  $19.9^{+0.6}_{-1.0}$ & 2.25$\pm$0.12 &  0.84$\pm$0.22 & $4.4^{+4.4}_{--4.4}$ &  69.4/55 \\
31531022 & 2013-11-27 07:07:59 &             2.0 &  $14.1^{+0.6}_{-1.1}$ & 2.27$\pm$0.16 &  0.77$\pm$0.33 & $4.4^{+4.4}_{--4.4}$ &  20.9/32 \\
31531023 & 2013-11-28 07:09:59 &             2.1 &  $13.1^{+0.7}_{-0.5}$ & 2.31$\pm$0.16 &  0.75$\pm$0.30 & $4.4^{+4.4}_{--4.4}$ &  20.5/30 \\
31531024 & 2013-11-29 07:11:59 &             2.0 &  $17.6^{+0.7}_{-0.9}$ & 2.18$\pm$0.14 &  0.57$\pm$0.27 & $4.4^{+4.4}_{--4.4}$ &  40.1/42 \\
31531025 & 2013-11-30 07:12:59 &             1.7 &  $23.5^{+0.9}_{-1.3}$ & 2.13$\pm$0.13 &  0.82$\pm$0.23 & $4.4^{+4.4}_{--4.4}$ &  43.5/46 \\
31531026 & 2013-12-01 08:50:59 &             1.9 &  $13.0^{+1.1}_{-1.1}$ & 2.49$\pm$0.16 &  0.05$\pm$0.31 & $4.4^{+4.4}_{--4.4}$ &  32.7/27 \\
31531027 & 2013-12-02 07:15:59 &             2.0 &  $15.1^{+1.1}_{-0.6}$ & 1.90$\pm$0.18 &  1.04$\pm$0.31 & $4.4^{+4.4}_{--4.4}$ &  38.6/36 \\
31531028 & 2013-12-03 07:17:59 &             1.9 &  $28.9^{+1.9}_{-1.1}$ & 2.28$\pm$0.12 &  0.51$\pm$0.25 & $4.4^{+4.4}_{--4.4}$ &  21.2/12 \\
32628005 & 2013-12-24 04:37:59 &             1.5 &   $9.5^{+0.7}_{-1.0}$ & 2.26$\pm$0.23 &  0.49$\pm$0.43 & $4.4^{+4.4}_{--4.4}$ &   6.7/16 \\
32628006 & 2013-12-25 04:32:59 &             1.5 &  $11.9^{+0.7}_{-1.0}$ & 2.00$\pm$0.20 &  1.11$\pm$0.39 & $4.4^{+4.4}_{--4.4}$ &  19.7/24 \\
32628007 & 2013-12-26 04:38:59 &             1.4 &  $15.9^{+0.9}_{-1.3}$ & 2.08$\pm$0.19 &  0.73$\pm$0.32 & $4.4^{+4.4}_{--4.4}$ &  31.4/27 \\
32628009 & 2013-12-28 04:35:59 &             1.5 &  $16.0^{+0.6}_{-1.0}$ & 2.09$\pm$0.16 &  0.93$\pm$0.27 & $4.4^{+4.4}_{--4.4}$ &  24.1/33 \\
32628010 & 2013-12-29 04:48:59 &             1.4 &   $6.8^{+0.6}_{-1.0}$ & 2.42$\pm$0.27 &  0.76$\pm$0.62 & $4.4^{+4.4}_{--4.4}$ &  10.7/11 \\
32628011 & 2013-12-30 04:32:59 &             1.4 &   $8.4^{+0.5}_{-0.6}$ & 2.34$\pm$0.22 &  1.20$\pm$0.47 & $4.4^{+4.4}_{--4.4}$ &  17.8/17 \\
32628012 & 2013-12-31 04:36:59 &             1.5 &   $7.2^{+0.5}_{-0.6}$ & 2.53$\pm$0.22 &  0.95$\pm$0.48 & $4.4^{+4.4}_{--4.4}$ &  15.2/15 \\
32628013 & 2014-01-01 04:36:59 &             1.5 &  $18.4^{+1.0}_{-1.1}$ & 2.20$\pm$0.15 &  0.60$\pm$0.26 & $4.4^{+4.4}_{--4.4}$ &  26.8/35 \\
32628014 & 2014-01-16 14:27:59 &             0.9 &  $18.4^{+2.0}_{-1.6}$ & 1.81$\pm$0.24 &  0.92$\pm$0.44 & $4.4^{+4.4}_{--4.4}$ &  12.7/18 \\
\enddata
%\tablecomments{}
\end{deluxetable*}

\bibliography{QiBibAll}

\allauthors

\end{document}